\documentclass[sigconf]{acmart}

\usepackage{xspace}
\usepackage{xcolor}
\usepackage[ruled,vlined,oldcommands,linesnumbered]{algorithm2e}
\usepackage{color, colortbl}
\usepackage{syntax}
\usepackage{amsmath}
\usepackage{booktabs}
\usepackage{listings}
\usepackage[T1]{fontenc}
\usepackage[scaled]{beramono}
\usepackage{hyperref}
\usepackage{paralist,enumitem}
\usepackage{tikz}
\usepackage{ragged2e}
\usepackage{wasysym}
\usepackage{subcaption}
\usepackage{amsfonts}
\usepackage{comment}

\usetikzlibrary{shapes,calc,arrows,positioning,fit}
\usepackage{pifont}
\usepackage{mathpartir}

\setcopyright{none}

\renewcommand\footnotetextcopyrightpermission[1]{} 
\begin{document}

\newcommand{\cmark}{\ding{51}}%
\newcommand{\gracom}[1]{\hfill \emph{(#1)}}
\newcommand{\sem}[1]{[\![#1]\!]}
\newcommand{\lang}{\textsc{SCL}\xspace}
\newcommand{\atp}{\textsc{FDL}\xspace}
\newcommand{\tl}{\textsc{TL}\xspace}
\newcommand{\heap}{\textit{M}}
\newcommand{\storage}{\textit{S}}
\newcommand{\store}{\textit{V}}
\newcommand{\balances}{\textit{Bal}}
\newcommand{\block}{\textit{B}}
\newcommand{\vars}{\mathcal{V}}
\newcommand{\preds}{\mathcal{P}}
\newcommand{\blockvars}{\mathcal{B}}
\newcommand{\transvars}{\mathcal{T}}
\newcommand{\offsets}{\mathcal{O}}
\newcommand{\addrs}{\mathcal{A}}
\newcommand{\vals}{\mathcal{C}}
\newcommand{\pc}{\textit{pc}}
\newcommand{\trans}{T}

\newcommand{\balance}[1]{\instruction{balance}(#1)}
\newcommand{\data}[0]{\instruction{data}}
\newcommand{\from}[0]{\instruction{caller}}
\newcommand{\fto}[0]{\instruction{To}}
\newcommand{\val}[0]{\instruction{callvalue}}
\newcommand{\assign}[2]{#1 := #2}
\newcommand{\assignpc}[3]{#1 :=^{#3} #2}
\newcommand{\call}[2]{#1\gets\instruction{call}(#2)}
\newcommand{\callpc}[3]{\instruction{call}(#3, #1, #2)}
\newcommand{\sha}[2]{#1\gets\instruction{Hash}(#2)}
\newcommand{\shapc}[3]{#1\gets\instruction{Hash}^{#3}(#2)}
\newcommand{\mload}[1]{\instruction{mload}(#1)}
\newcommand{\mloadpc}[2]{\instruction{mload}(#2,#1)}
\newcommand{\sload}[1]{\instruction{sload}(#1)}
\newcommand{\sloadpc}[2]{\instruction{sload}(#2,#1)}
\newcommand{\mstore}[2]{\instruction{mstore}[#1]\leftarrow#2}
\newcommand{\mstorepc}[3]{\instruction{mstore}(#3, #1, #2)}
\newcommand{\sstore}[2]{\instruction{sstore}[#1]\leftarrow#2}
\newcommand{\sstorepc}[3]{\instruction{sstore}(#3, #1, #2)}
\newcommand{\goto}[2]{\instruction{if}~#1~\instruction{goto}~#2}
\newcommand{\gotopc}[3]{\instruction{goto}(#3, #1, #2)}
\newcommand{\throw}[0]{\instruction{throw}}
\newcommand{\throwpc}[1]{\instruction{throw}(#1)}
\newcommand{\sstop}[0]{\instruction{stop}\xspace}
\newcommand{\sstoppc}[1]{\instruction{stop}(#1)}
\newcommand{\sstate}[1]{(#1)}
\newcommand{\states}[0]{\Sigma}
\newcommand{\config}[1]{\langle #1 \rangle}

\newcommand{\connector}[1]{\text{#1}}
\newcommand{\myforall}{\relation{all}\ }
\newcommand{\myexists}{\relation{some}\ }
\newcommand{\myneg}{\neg}
\newcommand{\myand}{\wedge}
\newcommand{\myor}{\vee}
\newcommand{\myimplies}{\Rightarrow }

\newcommand{\maydependon}{\relation{MayDepOn}}
\newcommand{\determinedby}{\relation{DetBy}}
\newcommand{\eq}{\relation{Eq}}
\newcommand{\mustdependon}{\relation{MustDepOn}}
\newcommand{\mustfollow}{\relation{MustFollow}}
\newcommand{\mayfollow}{\relation{MayFollow}}
\newcommand{\follow}{\relation{Follow}}

\newcommand{\Dana}[1]{{\color{purple} {\bf Dana:} #1}}
\setlength{\grammarparsep}{2pt plus 1pt minus 1pt}

\tikzstyle{block} = [rectangle, draw, fill=white!5,
    align=left,
    rounded corners,
    ]
\tikzstyle{whiteblock} = [rectangle, draw=white, fill=white!5,
    align=center,
    rounded corners,
    ]

\renewcommand\labelitemi{--}

\tikzstyle{line} = [draw, -latex',align=left]
\newcommand{\st}[1]{#1}
\newcommand{\dsl}{TBD\xspace} 
\newcommand{\evm}{\textsc{EVM}\xspace}
\newcommand{\para}[1]{\vspace{6pt}\noindent{\bf #1.}\hspace{4pt}}
\newcommand{\tool}{\textsc{Securify}\xspace}
\definecolor{darkred}{rgb}{0.75,0,0}
\definecolor{darkgreen}{rgb}{0,0.5,0}
\definecolor{darkpurple}{rgb}{0.58,0,0.333}
\newcommand{\todo}[1]{{\color{blue} [#1] - PT}}
\newcommand{\terminal}[1]{{\color{blue}{#1}}}
\newcommand{\ad}[1]{ {\color{purple} \bf AD: #1} }
\newcommand{\code}[1]{{\small\texttt{\detokenize{#1}}}}
\newcommand{\oyente}[0]{\textsc{Oyente}\xspace}
\newcommand{\relation}[1]{{\textit{\color{darkpurple}  #1}}}
\newcommand{\labelc}[1]{{{\color{darkpurple}  #1}}}
\newcommand{\instruction}[1]{{\text{\color{blue}  #1}}}
\newcommand{\wildcard}{\_}
\newcommand{\bodyf}[1]{\text{ #1}}

\definecolor{mygreen}{rgb}{0.749,1,0.776}
\definecolor{myblue}{rgb}{0.8413,0.878,1}
\definecolor{myred}{rgb}{1,0.647,0.647}

\newcommand*\LSTfont{\footnotesize}

\lstset{
        language=Java,
        xleftmargin=14pt,
        numbers=left,
        basicstyle=\LSTfont,
        numberstyle=\color{gray}\LSTfont,
        keywordstyle=\color{blue}\LSTfont,
        commentstyle=\color{darkgreen},
        morekeywords={*,Push,Call,Gas,Timestamp,Number,GasLimit,GasPrice,Balance,BlockHash,Difficulty,MessageField,MessageCaller, Add, Mul, Sub, Address, Caller, MStore, MLoad, SStore, SLoad, Jump, JumpI,JumpDest, Stop, Suicide,CallDataSize,CallDataLoad, CallValue,LT,EQ,NOT,IsZero,SHA3,SHA256,RIPEMD160,CoinBase,Swap1,Dup1,Log0,Pop,Push1,Push2,Push32,MStore8,contract,uint,function,call,bool,public},
        numbersep=5pt,
        frame=none,
        firstnumber=auto,
        escapeinside={(*@}{@*)},
        captionpos=b,
}

\title{\tool: Practical Security Analysis of Smart Contracts} 

\author{Petar Tsankov}
\affiliation{%
  \institution{ETH Zurich}
}
\email{petar.tsankov@inf.ethz.ch}

\author{Andrei Dan}
\affiliation{%
  \institution{ETH Zurich}
}
\email{andrei.dan@inf.ethz.ch}

\author{Dana Drachsler-Cohen}
\affiliation{%
  \institution{ETH Zurich}
}
\email{dana.drachsler@inf.ethz.ch}

\author{Arthur Gervais}
\authornote{Work done while at ETH Zurich}
\affiliation{%
  \institution{Imperial College London}
}
\email{a.gervais@imperial.ac.uk}

\author{Florian B\"unzli}
\affiliation{%
  \institution{ETH Zurich}
}
\email{fbuenzli@student.ethz.ch}

\author{Martin Vechev}
\affiliation{%
  \institution{ETH Zurich}
}
\email{martin.vechev@inf.ethz.ch}


\renewcommand{\shortauthors}{P. Tsankov, A. Dan, D. Drachsler-Cohen, A. Gervais, F. B\"unzli, M. Vechev}

\begin{abstract}
Permissionless blockchains allow the execution of arbitrary programs (called {\em smart contracts}), enabling mutually untrusted entities to interact without relying on trusted third parties. Despite their potential, repeated security concerns have shaken the trust in handling billions of USD by smart contracts.

To address this problem, we present \tool, a security analyzer for Ethereum smart contracts that is scalable, fully automated, and able to prove contract behaviors as safe/unsafe with respect to a given property. \tool's analysis consists of two steps. First, it symbolically analyzes the contract's dependency graph to extract precise semantic information from the code. Then, it checks compliance and violation patterns that capture sufficient conditions for proving if a property holds or not. To enable extensibility, all patterns are specified in a designated domain-specific language.

\tool is publicly released, it has analyzed $>18K$ contracts submitted by its users, and is regularly used to conduct security audits by experts. We present an extensive evaluation of \tool over real-world Ethereum smart contracts and demonstrate that it can effectively prove the correctness of smart contracts and discover critical violations.
\end{abstract}


\maketitle

\definecolor{warning}{RGB}{247, 147, 30}
\definecolor{violation}{RGB}{237, 28, 36}
\definecolor{compliance}{RGB}{0, 104, 55}
\newcommand{\violation}{{\color{violation}$\Diamondblack$}\xspace}
\newcommand{\warning}{{\large\color{warning}$\blacktriangle$}\xspace}
\newcommand{\compliance}{{\small\color{compliance}$\blacksquare$}\xspace}

\section{Introduction} \label{sec:intro}
Blockchain platforms, such as Nakamoto's Bitcoin~\cite{nakamoto2008bitcoin}, enable the trade of crypto-currencies between mutually mistrusting parties. To eliminate the need for trust, Nakomoto designed a peer-to-peer network that enables its peers to agree on the trading transactions. Buterin~\cite{whitepaper} identified the applicability of decentralized computation beyond trading, and  designed the Ethereum blockchain which supports the execution of programs, called smart contracts, written in Turing-complete languages. Smart contracts have shown to be applicable in many domains including financial industry~\cite{blockchaininsurance}, public sector~\cite{recordkeeping} and cross-industry~\cite{ethlance}.

The increased adoption of smart contracts demands strong security guarantees. Unfortunately, it is challenging to create smart contracts that are free of security bugs. As a consequence, critical vulnerabilities in smart contracts are discovered and exploited every few months~\cite{kingofether,thedao,etherdice,bylica17,paritybug,paritybug2}. In turn, these exploits have led to losses reaching millions worth of USD in the past few years: $150$M were stolen from the popular DAO contract in June $2016$~\cite{thedao}, $30$M were stolen from the widely-used Parity multi-signature wallet in July $2017$~\cite{paritybug}, and few months later $280$M were frozen due to a bug in the very same wallet~\cite{parity2}. It is apparent that effective security checkers for smart contracts are urgently needed.

\para{Key Challenges}
The main challenge in creating an effective security analyzer for smart contracts is the Turing-completeness of the programming language, which renders automated verification of arbitrary properties undecidable. To address this issue, current automated solutions tend to rely on fairly generic testing and symbolic execution methods (e.g., Oyente~\cite{luu2016making} and Mythril~\cite{mythril}). While useful in some settings, these approaches come with several drawbacks: {\em (i)} they can miss critical violations (due to under-approximation), {\em (ii)} yet, can also produce false positives (due to imprecise modeling of domain-specific elements~\cite{GrishchenkoMS18}), and {\em (iii)} they can fail to achieve sufficient code coverage on realistic contracts (Oyente achieves only $20.2\%$ coverage on the popular Parity wallet~\cite{walletlibrary}). Overall, these drawbacks place a significant burden on their users, who must inspect all reports for false alarms and worry about unreported vulnerabilities. Indeed, many security properties for smart contracts are inherently difficult to reason about directly. 
A viable path to addressing these challenges is building an automated verifier that targets important domain-specific properties~\cite{best-practices}. For example, recent work~\cite{GrossmanAGMRSZ18} focuses solely on identifying reentrancy issues in smart contracts~\cite{reentrancy-blog}.

\para{Domain-Specific Insight}
A key observation of this work is that it is often possible to devise precise patterns expressed on the contract's data-flow graph in a way where a match of the pattern implies either a violation or satisfaction of the original security property.
For example, $90.9\%$ of all calls in Ethereum smart contracts can be proved free of the infamous DAO bug~\cite{thedao} by matching a pattern stating that calls are not followed by writes to storage.
The reason why it is possible to establish such a correspondence is that violations of the original property in real-world contracts tend to often violate a much simpler property (captured by the pattern).
Indeed, in terms of verification, a key benefit in working with patterns, instead of with their corresponding property, is that patterns are substantially more amenable to automated reasoning.

\para{\tool: Domain-specific Verifier}
Based on the above insight, we developed \tool, a lightweight and scalable security verifier for Ethereum smart contracts.
The key technical idea is to define two kinds of patterns that mirror a given security property: {\em (i)} compliance patterns, which imply the satisfaction of the property, and {\em (ii)} violation patterns, which imply its negation.
To check these patterns, \tool symbolically encodes the dependence graph of the contract in stratified Datalog~\cite{Ullman:1988} and leverages off-the-shelf scalable Datalog solvers to efficiently (typically within seconds) analyze the code.
To ensure extensibility, all patterns are expressed in a designated domain-specific language (DSL).

\begin{figure}
\includegraphics[width=1\columnwidth]{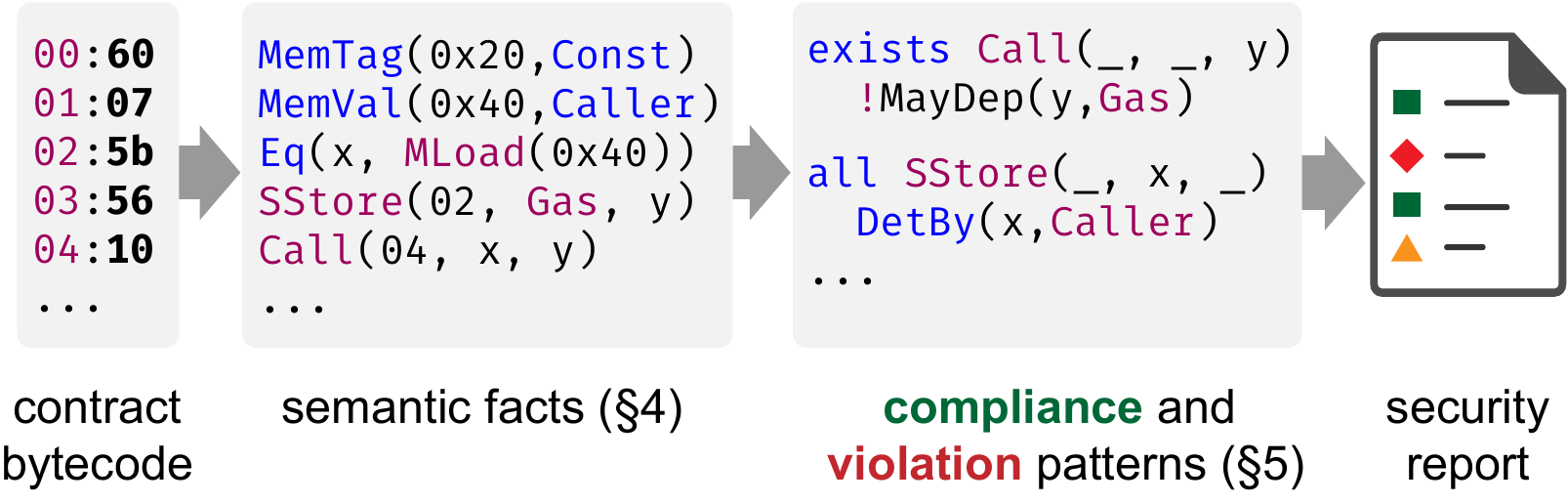}
\caption{\tool's approach is based on automatic inference of semantic program facts followed by checking of compliance and violation security patterns over these facts.}
\label{fig:intro-flow}
\end{figure}

\begin{figure}
\includegraphics{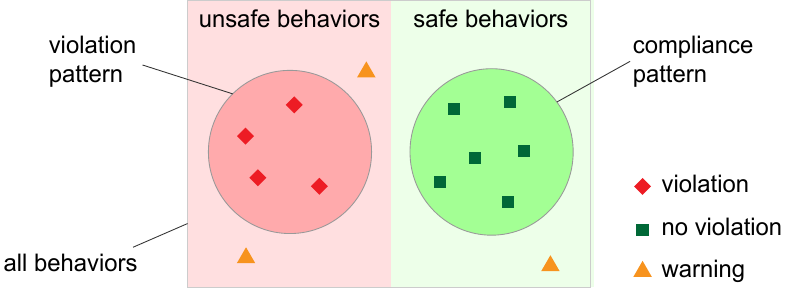}
\caption{
\tool uses compliance and violation patterns to guarantee that certain behaviors are safe and, respectively, unsafe. The remaining behaviors are reported as warnings (to avoid missing errors).}
\label{fig:intro-patterns}
\end{figure}

In Fig.~\ref{fig:intro-flow}, we illustrate the analysis flow of \tool. Starting with the contract's bytecode (or source code, which can be compiled to bytecode), \tool derives semantic facts inferred by analyzing the contract's dependency graph and uses these facts to check a set of compliance and violation patterns. Based on the outcome of these checks, \tool classifies all contract behaviors into violations~(\violation), warnings~(\warning), and compliant~(\compliance), as abstractly illustrated in Fig.~\ref{fig:intro-patterns}. Here, the large box depicts all contract behaviors, partitioned into safe (which satisfy the property) and unsafe ones (which violate it). \tool reports as violations (\violation) all behaviors matching the violation pattern, and as warnings (\warning) all remaining behaviors not matched by the compliance pattern.


\para{Reduced Manual Effort}
Compared to existing symbolic analyzers for smart contracts, \tool reduces the required effort to inspect reports in two ways. First, existing analyzers do not report definite violations (they conflate  \violation and  \warning), and thus require users to manually classify {\em all} reported vulnerabilities into true positives (found in the \colorbox{red!14}{red box}) or false positives (found in the \colorbox{green!12}{green box}). In contrast, \tool automatically classifies behaviors guaranteed to be violations (marked with \violation). Hence, the user only needs to manually classify the warnings (\warning) as true or false positives. 

As we show in our evaluation, the approach of using both violation and compliance patterns reduces the warnings a user needs to inspect manually by $65.9\%$, and even up to $99.4\%$ for some properties. Second, existing analyzers fail to report unsafe behaviors (sometimes up to $72.9\%$), meaning users may have to manually inspect portions of the code that are not covered by the analyzer. In contrast, \tool reports all unsafe behaviors.

\para{Auditing Smart Contracts}
\tool is publicly available at \url{https://securify.ch} and has analyzed $> 18K$ contracts submitted by its users. Over the last year, we have also extensively used \tool to perform 38 detailed commercial audits of smart contracts (other auditors have also used \tool), iteratively improving the approach and adding more patterns. Indeed, the design and implementation of \tool have greatly benefited from this experience. 

In terms of the actual audit process, our approach (and we believe that of other auditors) has been to run all available tools and then to manually inspect the reported vulnerabilities so to assess their severity. For instance, while \tool covers a number of important properties (the full version supports $18$ properties), symbolic execution tools have better support for numerical properties (e.g., overflow). Our finding was that \tool was particularly helpful in auditing larger contracts, which are challenging to inspect with existing solutions for the reasons listed earlier. Overall, we believe \tool is a pragmatic and valuable point in the space of analyzing smart contracts due to its careful balance of scalability, guarantees, and precision. 

\para{Main Contributions}
To summarize, our main contributions are:	
\begin{itemize}[nosep,nolistsep]
	\item A decompiler that symbolically encodes the dependency graph of Ethereum contracts in Datalog (Section~\ref{sec:analysis}).
	\item A set of compliance and violation security patterns that capture sufficient conditions to prove and disprove practical security properties (Section~\ref{sec:patterns}).
	\item An end-to-end implementation, called \tool, which fully automates the analysis of contracts (Section~\ref{sec:implementation}).
	 \item An extensive evaluation over existing Ethereum smart contracts showing that \tool can effectively prove the correctness of contracts and discover violations (Section~\ref{sec:evaluation}).
\end{itemize}

\section{Motivating Examples}\label{sec:motivation}
In this section, we motivate the problem we address through 
two real-world security issues that affected $\approx 200$ millions worth of USD in $2017$. We describe the underlying security properties and the challenges involved in proving whether a contract satisfies/violates them. We also describe how \tool discovers both vulnerabilities with appropriate violation patterns.

\subsection{Stealing Ether}

\label{sec:stealing-ownership}

\begin{figure}[t]
\centering
\includegraphics[width=0.94\columnwidth]{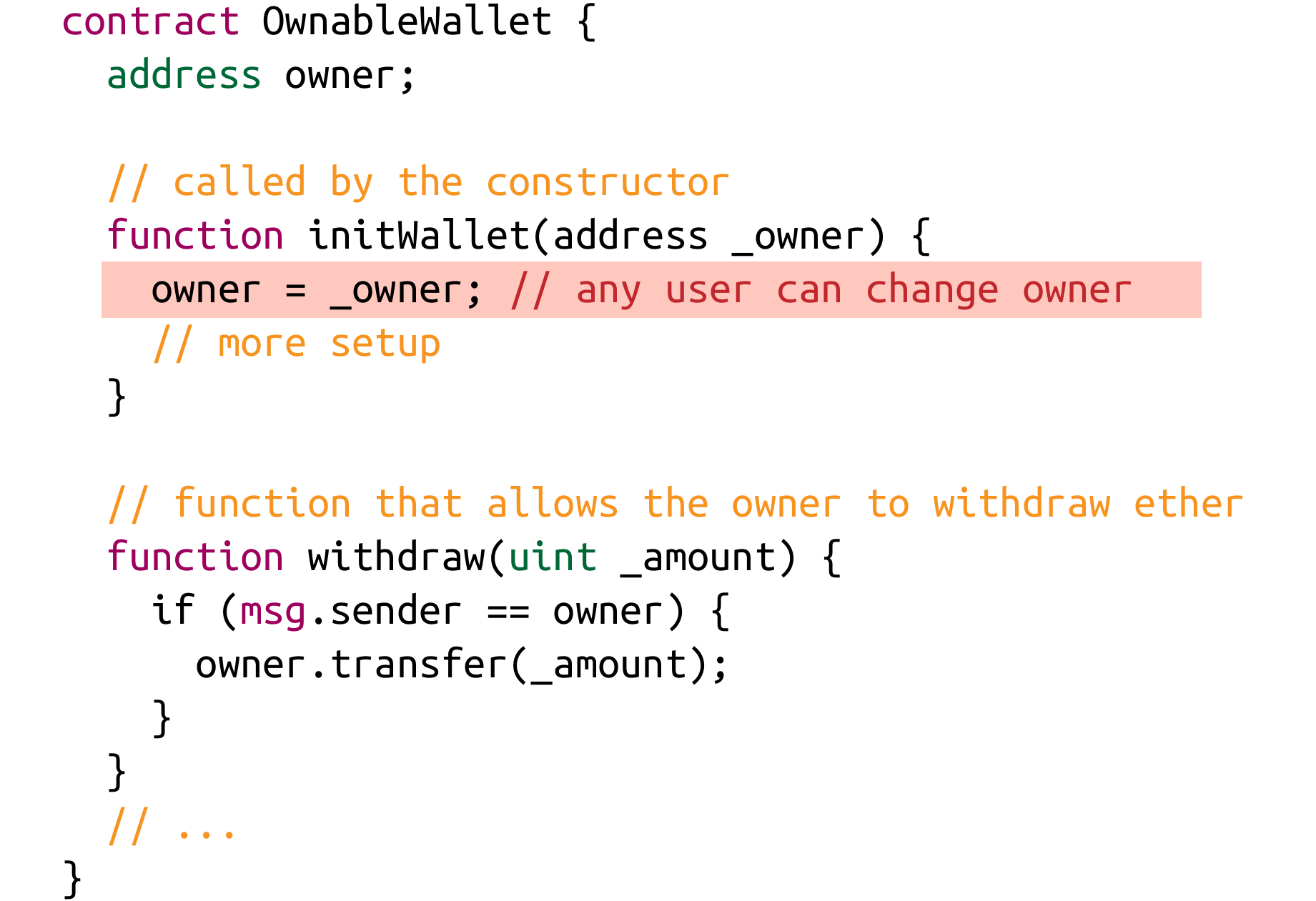}
\caption{A vulnerable wallet that allows any user to withdraw all ether stored in it.}
\label{fig:bug1}
\end{figure}

In Fig.~\ref{fig:bug1}, we show an implementation of a wallet. The code is written in Solidity~\cite{solidity-lang}, a popular high-level language for writing Ethereum smart contracts. We remark that this wallet is a simplified version of Parity's multi-signature wallet, which
allowed an attacker to steal $30$ million worth of USD in July 2017.

The wallet has a field \code{owner}, which stores the address of the wallet's owner.
Further, the contract has a function \code{initWallet}, which takes as argument an address \code{_owner} and initializes the field \code{owner} with it.
This function is called by the constructor (not shown in Fig.~\ref{fig:bug1}), and was assumed not to be accessible otherwise~\cite{paritybug}.
Finally, the contract has a function \code{withdraw}, which takes as argument an unsigned integer \code{_amount}. The function checks if the transaction sender's address (returned by \code{msg.sender}) equals that of the contract's owner (stored in the field \code{owner}). If this check succeeds, it transfers \code{_amount} ether to the owner with the statement \code{owner.transfer(_amount)}; otherwise, no ether is transferred. The \code{withdraw} function ensures that only the owner can withdraw ether from the wallet.

\para{Attack}
The wallet shown in Fig.~\ref{fig:bug1} has a critical security flaw: any user could actually call the \code{initWallet} function and store an arbitrary address in the field \code{owner}.
An attacker can, therefore, steal all ether stored in the wallet in two steps. First, the attacker calls the function \code{initWallet}, passing her own address as argument. Second, the attacker calls the function \code{withdraw}, passing as argument the amount of ether stored in the wallet. We remark that in the attack on Parity's wallet, to perform the first step the attacker exploits a fallback mechanism to call the \code{initWallet} function; we omit these details for simplicity and refer the reader to~\cite{paritybug} for details on the actual attack.

\para{Security Property}
The underlying security problem that allowed the attacker to steal ether is that the security-critical field \code{owner} is universally writable by {\em any} Ethereum user. 
This security issue mirrors a more general property stipulating that the write to the \code{owner} field is restricted, in the sense that not all users can make a transaction that writes to this field. 
To show that this property is satisfied, we need to demonstrate that some user cannot send a transaction that modifies the \code{owner} field. Conversely, to show a violation, we need to prove that all users can send a transaction that modifies the \code{owner} field. Proving both satisfaction and demonstrating violations of this property is nontrivial due to the enormous space of possible users and transactions that they can make.

\para{Detection}
To discover this security issue, \tool provides a violation pattern that is matched if the execution of the assignment \code{owner = _owner}, highlighted in \colorbox{myred}{red} in Fig.~\ref{fig:bug1}, does not depend on the value returned by the \instruction{caller} instruction (which returns the address of the transaction sender).
To check this pattern, \tool infers data- and control-flow dependencies by analyzing the contract's dependency graph; cf.~\cite{Johnson:2015:EES:2737924.2737957}.
Here, \tool infers that the assignment \code{owner = _owner} does not depend on the \instruction{caller} instruction, which implies that the assignment is reachable by any user.
In Section~\ref{sec:overview}, we provide more details on this violation pattern and further details on how \tool uses it to detect the vulnerability.

We remark that some symbolic checkers perform imprecise checks of similar properties, which result in both false positives and false negatives.
For instance, as we show in Fig.~\ref{fig:compare} of our evaluation later, Mythril~\cite{mythril} has about $65\%$ false negatives when checking a similar property stipulating that not all users may trigger a particular ether transfer.

\begin{figure}[t]
\centering
\includegraphics[width=0.94\columnwidth]{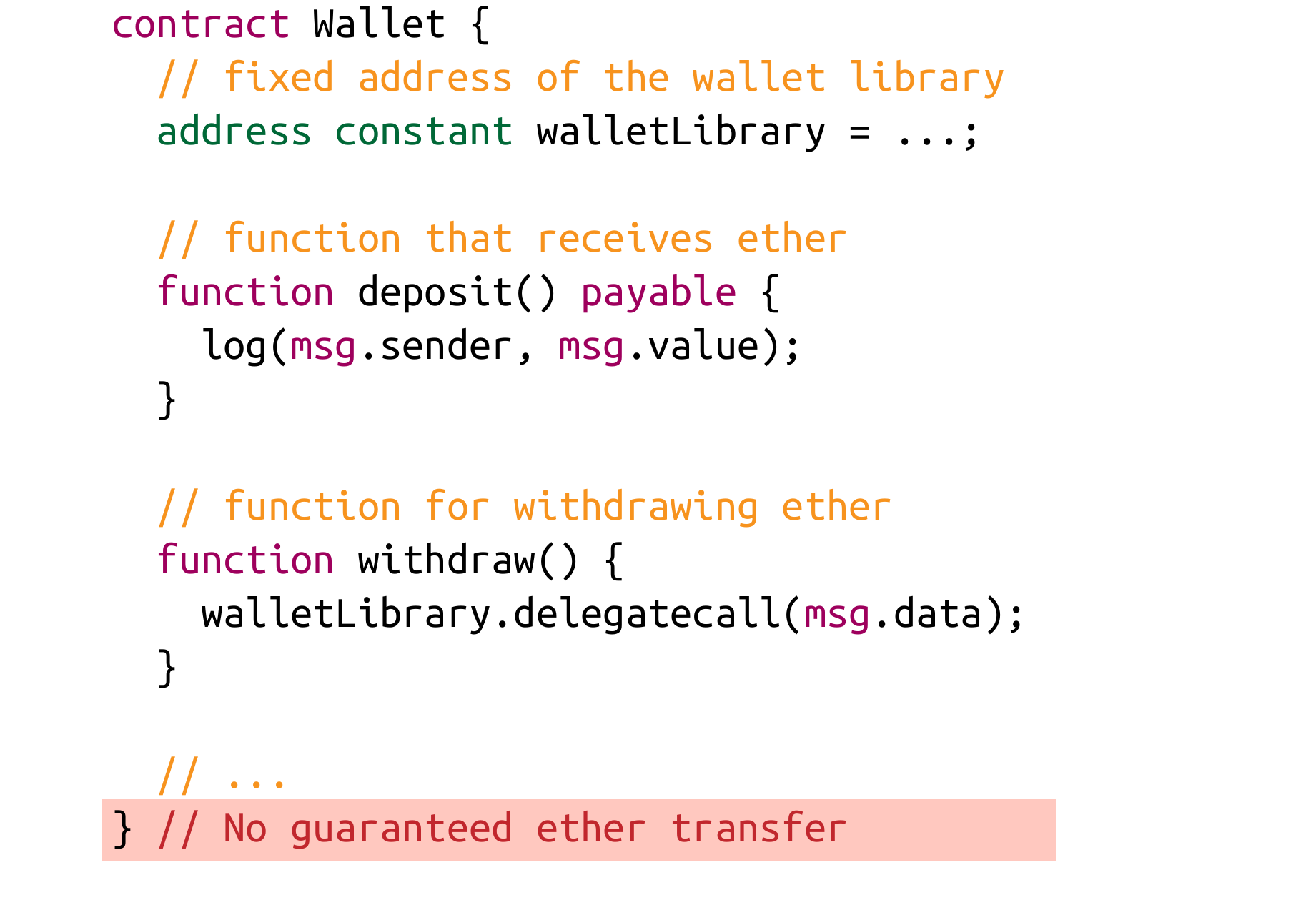}
\caption{A wallet that delegates functionality to a library contract \code{walletLibrary}.}
\label{fig:bug2}
\end{figure}

\begin{figure*}[t!]
	\centering
	\includegraphics[width=0.9\textwidth]{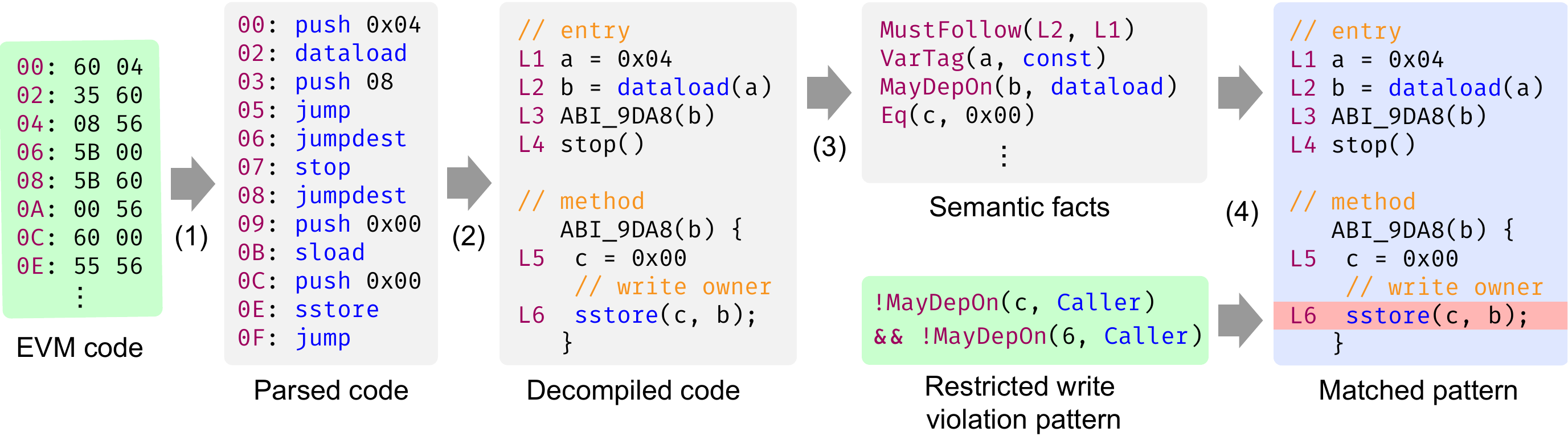}
	\caption{
		High-level flow illustrating how \tool finds the unrestricted write to the \code{owner} field of the contract from Fig.~\ref{fig:bug1}.
		The input (EVM bytecode and security patterns) is highlighted in \colorbox{mygreen}{green}, the output (in our example, a violated instruction) is highlighted in \colorbox{myred}{red}, and \colorbox{gray!20}{gray} boxes represent intermediate analysis artifacts. \tool proceeds in three steps:
		(1) it decompiles the contract's EVM bytecode into a static-single assignment form, (2) it infers semantics facts about the contract, and (3) it matches the violation pattern of the restricted write property on the \code{sstore} instruction that writes to the  \code{owner} field.}
	\label{fig:overview}
\end{figure*}

\subsection{Frozen Funds}
\label{sec:frozen}

In Fig.~\ref{fig:bug2}, we show a wallet implementation which suffers from a security issue that froze millions worth of USD in November $2017$.
This wallet has a field, \code{walletLibrary}, which stores the address of a contract implementing common wallet functionality.
Further, it has a function \code{deposit}, marked as \relation{payable}, which means users can send ether to the contract by calling this function. The function \code{deposit} logs the amount of ether (identified by \code{msg.value}) sent by the transaction sender (identified by \code{msg.sender}).
Finally, the contract has a function \code{withdraw}, which delegates all calls to the wallet library. That is, the statement \code{walletLibrary.delegatecall(msg.data)} results in executing the \code{withdraw} method of the wallet library in the context of the current wallet.

\para{Attack}
Ethereum contracts can be removed from the blockchain using a designated \code{kill} instruction.
If an attacker can remove the wallet library from the blockchain, then the funds in the wallet cannot be extracted from the wallet. This is because the wallet relies on the library smart contract to withdraw ether. In November $2017$, a popular wallet library was removed from the blockchain, effectively freezing $\approx 280$ million worth of USD~\cite{paritybug2}.


\para{Security Property}
The underlying security problem with this wallet is that it allows users to deposit ether, but it cannot guarantee that the ether can be transferred out of the contract, since the transfer depends on a library.
To discover that the wallet has this problem, we must prove two facts: {\em (i)} users can deposit ether and {\em (ii)} the contract has no ether transfer instructions (i.e., \instruction{call}) with non-zero amount of ether. Note that if the contract only transfers out ether through libraries, the second requirement is met. 

\para{Detection}
To discover this vulnerability, \tool's violation pattern checks the conjunction of two facts. First, to prove that users can deposit ether, \tool checks whether there is a \instruction{stop} instruction whose execution does not depend on the ether transferred being zero. Assuming that the \instruction{stop} instruction is reachable for some transaction, this implies that a user can reach it \emph{with a positive ether amount}, resulting in a deposit of ether to the contract. Second, \tool checks whether for all \instruction{call} instructions, the amount of ether extracted from the contract is zero. The conjunction of these two facts implies that ether can be locked in the contract.

\section{The \tool System}\label{sec:overview}
In the previous section, we illustrated that while security issues in smart contracts are complex, they can be often captured with semantic facts inferred from the code.
In this section, we describe the \tool system, which builds on this idea to prove and disprove security properties of smart contracts.
We accompany this section with the example of how \tool detects the unrestricted write to the \code{owner} field in the wallet contract (Fig.~\ref{fig:bug1}). Fig.~\ref{fig:overview} summarizes the main steps.

\para{Inputs to \tool}
The input to \tool is the EVM bytecode of a contract and a set of security patterns, specified in our designated domain-specific language (DSL). \tool can also take as input contracts written in Solidity (not shown in Fig.~\ref{fig:overview}), which are compiled to EVM bytecode before proceeding with the analysis. 
There are two kinds of security patterns:
compliance and violation patterns, which capture sufficient conditions to ensure that a contract satisfies and, respectively, violates a given security property.

Fig.~\ref{fig:overview} illustrates the input to \tool in \colorbox{mygreen}{green} boxes, which show part of the EVM bytecode of the wallet contract (only the part necessary to illustrate the vulnerability) and the 
violation pattern of the \emph{restricted write} property. Intuitively, the pattern is matched if there is a write that is \emph{not} restricted.

To discover the unrestricted write in the contract, \tool proceeds with the following three steps.

\para{Step 1: Decompiling EVM Bytecode}
\tool first transforms the EVM bytecode provided as input into a stackless representation in static-single assignment form (SSA).
For example, in Fig.~\ref{fig:overview},
for the stack expression \code{push 0x04}, \tool introduces a local variable \code{a} and an assignment statement \code{a = 4}. In addition to removing the stack, \tool identifies methods. For example, the method \code{ABI_9DA8}, shown in Fig.~\ref{fig:overview}, corresponds to the \code{initOwner} method of the wallet contract, shown in Fig.~\ref{fig:bug1}.
After decompilation, \tool performs partial evaluation to resolve memory and storage offsets, jump destinations, all of which are important for precisely analyzing the code statically. We describe these optimizations in Section~\ref{sec:implementation}.

\para{Step 2: Inferring Semantic Facts}
After decompilation,
\tool analyzes the contract to infer semantic facts, including data- and control-flow dependencies, which hold over all behaviors of the contract. For example, the fact $\relation{MayDepOn}(\code{b}, \instruction{dataload})$, shown in Fig.~\ref{fig:overview}, captures that the value of variable \code{b} may depend on the value returned by the instruction \instruction{dataload}. Further, the fact $\relation{Eq}(\code{c}, 0)$ captures that variable \code{c} equals the constant~$0$.

\tool's derivation of semantic facts is specified declaratively in stratified Datalog and is fully automated using existing scalable engines~\cite{souffle}. Key benefits of the declarative approach are: \textit{(i)}~inference rules concisely capture abstract reasoning about different components (e.g., contract storage), \textit{(ii)}~more facts and inference rules can be easily added, and \textit{(iii)}~inference rules are specified in a modular way (e.g., memory analysis is specified independently of contract storage analysis). We list the semantic facts that \tool derives, along with the inference rules, in Section~\ref{sec:analysis}.

\para{Step 3: Checking Security Patterns}
After obtaining the semantics facts, \tool checks the set of compliance and violation security patterns, given as input.
These patterns are written in a specialized domain-specific language (DSL), which enables security experts to extend our built-in set of patterns with their customized patterns.
Our DSL is a fragment of logical formulas over the semantic facts inferred by \tool.
To detect the vulnerability in the contract of Fig.~\ref{fig:bug1}, \tool matches the violation pattern (given as input) on the $\instruction{sstore}(\code{c},\code{b})$ instruction at label \labelc{l6} in Fig.~\ref{fig:overview}. In $\instruction{sstore}(\code{c},\code{b})$, \code{c} is the storage offset of the \code{owner} field, and \code{b} is the value to store.
The violation pattern matches if there exists some \instruction{sstore} instruction for which both the storage offset, denoted~$X$, and the execution of this instruction, identified by its label $L$, do not depend on the result of the \instruction{caller} instruction in \emph{any possible execution} of the contract.
Since the instruction \instruction{caller} retrieves the address of the transaction sender, matching this pattern implies that
 any user can reach this \instruction{sstore} and change the value of \code{owner}.
In our DSL, where negation is encoded by $\myneg$ and conjunction by $\myand$, this pattern is encoded as:
\[\myexists \sstorepc{X}{\_}{L}.\ \myneg \maydependon(X, \instruction{caller}) \myand \myneg \maydependon(L, \instruction{caller})\]

\tool's DSL is important for extensibility: adding new security patterns amounts to specifying them in this DSL.
To illustrate the expressiveness of the DSL, in Section~\ref{sec:patterns}, we present a range of  security patterns for important properties, such as restricted writes, exception handling, ether liquidity, input validation, and others.

We remark that contract-specific patterns are sometimes added by security experts while conducting security audits. For example, it is often required to check for the absence of undesirable dependencies, such as: only the owner can modify certain values in the storage, or to ensure that the result of a specific arithmetic expression does not depend on the division instruction (which may cause undesirable integer rounding effects). We illustrate how such contract-specific patterns are specified in the DSL in Section~\ref{sec:patterns}.

\para{Output of \tool} 
For any match of a violation pattern, \tool outputs the instruction that caused the pattern to match. In our example, it highlights the instruction $\instruction{sstore}(\code{c},\code{b})$. We remark that the offset of this instruction can be easily mapped to its corresponding line in the Solidity code, if the source code is provided.
Further, for any property for which neither the violation nor the compliance pattern is matched, \tool outputs a warning, indicating that it failed to prove or disprove the property.

\para{Limitations}
We briefly summarize several limitations of \tool. First, the current version of \tool cannot reason about numerical properties, such as overflows. To address this limitation, we plan to extend \tool with numerical analysis (e.g., using ELINA~\cite{Singh:2017:FPA:3009837.3009885}), which would not only improve the precision of \tool but also enable the checking of numerical properties. 

Second, \tool does not reason about reachability, and assumes that all instructions in the contract are reachable. This assumption is necessary to establish a formal correspondence between the security properties supported by \tool and the patterns used to prove and disprove them. For instance, in our example, \tool assumes that the matched $\instruction{sstore}$ instruction is reachable by some execution (otherwise, there is no violation). 

Finally, the properties we consider capture violations that can often, but not always, be exploited by attackers. For example, there are fields in the contract that must be universally writable by all users. To address this, security experts can write contract-specific patterns in \tool's DSL (e.g., to specify which fields are sensitive).

\section{Semantic Facts}\label{sec:analysis}
In this section, we present the automated inference of control- and data-flow dependencies that \tool employs. The facts inferred in this process are called {\em semantic facts} and are later used for checking security properties.  
We begin with the background necessary for understanding this analysis: the EVM instruction set and stratified Datalog. We then introduce the semantic facts derived by \tool and the declarative inference rules, specified in stratified Datalog, used to derive them.

\subsection{Background}

In this section, we provide the necessary background.

\subsubsection{Ethereum Virtual Machine (EVM)}\label{sec:background}

Smart contracts are executed on a \emph{blockchain}. A contract executes when a user submits a \emph{transaction} that specifies the contract, the method to run, and the method's arguments (if any). When the transaction is processed, it is added to a new block, which is appended to the blockchain.
Contracts can access a volatile memory and non-volatile storage.
The EVM instruction set (over which contracts are written) supports a few dozen opcodes. \tool handles all EVM opcodes; we present the most relevant ones below. Note that many of the opcodes (such as \instruction{push}, \instruction{dup}, etc.) are eliminated when \tool decompiles the EVM bytecode to its stackless representation. The relevant instructions are:

\begin{compactitem}
\item Arithmetic operations and comparisons: e.g., \instruction{add}, \instruction{mul}, \instruction{lt}, \instruction{eq}. In the rest of the paper, we write \instruction{op} to denote any of these operations.
\item Cryptographic hash functions: e.g., \instruction{sha3}.
\item Environmental information: e.g., \instruction{balance} returns the balance of
a contract, \instruction{caller} is the identity of the transaction sender, \instruction{callvalue} is the amount of ether specified to be transferred by the transaction.
\item Block information: e.g., \instruction{number}, \instruction{timestamp}, \instruction{gaslimit}.
\item Memory and storage operations:
\instruction{mload},  \instruction{mstore},  \instruction{sstore},  \instruction{sload} 
load/store data from the memory/contract storage.
\item System operations: e.g., \instruction{call}, which transfers ether, and takes two arguments: receiver address and amount of ether to transfer (in fact, \instruction{call} takes seven arguments; we consider here only those that are relevant for the rest of the paper).
\item Control-flow instructions: e.g., \instruction{goto}, which encodes conditional jumps across instructions.
\end{compactitem}

For the complete set of instructions, along with their formal description, we refer the reader to~\cite{wood2014ethereum}.

\newcommand{\getvars}[1]{\textit{vars}(#1)}
\newcommand{\edb}[1]{\textit{edb}(#1)}
\newcommand{\idb}[1]{\textit{idb}(#1)}

\subsubsection{Stratified Datalog}
Stratified Datalog is a declarative logic language, which enables to write \emph{facts} (predicates) and \emph{rules} to infer facts.
We next briefly overview its syntax and semantics.

\para{Syntax}
We present Datalog's syntax in Fig.~\ref{fig:datalog}.
A Datalog program consists of one or more rules, denoted $\overline{r}$.
A rule $r$ consists of a head $a$, and a body, $\overline{l}$, consisting of literals, separated by commas.
The head, also called an atom, is a predicate over zero or more terms, denoted~$\overline{t}$, comma-separated.
A literal $l$ is a predicate or its negation. 
  As a convention, we write Datalog variables in upper case and constants in lower case.
A Datalog program is {\em well-formed} if for any rule $a\Leftarrow \overline{l}$, we have $\getvars{a}\subseteq \getvars{\overline{l}}$, where $\getvars{\overline{l}}$ returns the set of variables in $\overline{l}$.

A Datalog program~$P$ is {\em stratified} if its rules can be partitioned into strata $P_1, \ldots, P_n$ such that if a predicate $p$ occurs in a positive (negative) literal in the body of a rule in $P_i$, then all rules with $p$ in their heads are in a stratum $P_j$ with $j \leq i$ ($j< i$). Stratification ensures that predicates that appear in negative literals are fully defined in lower strata.

\begin{figure}[t]
	\centering
	\setlength{\tabcolsep}{2pt}
	\renewcommand{\arraystretch}{1}
	\begin{tabular}{rrclrrclrrcl}
		{\em (Program)} & $P$ & $::=$ & $\overline{r}$ \hspace{30pt} 	& {\em (Predicates)} &$p, q$ & $\in$ & $\preds$\\	
		{\em (Rule)} & $r$ & $::=$ & $a \Leftarrow \overline{l}$		& {\em (Term)} & $t$ & $\in$ & $\vars \cup \vals$\\
		{\em (Atom)} & $a$ & $::=$ & $p(\overline{t})$ 					& {\em (Datalog variables)} & $X, Y$ & $\in $ & $\vars$\\
		{\em (Literal)} & $l$ & $::=$ & $a\mid \neg a$					& {\em (Constants)} & $x, y$ & $\in$ & $\vals$\\
	\end{tabular}\\
	\caption{Syntax of stratified Datalog.}
	\label{fig:datalog}
\end{figure}

\para{Semantics}
Let $\mathcal{A} = \{ p(\overline{t})\mid \overline{t}\subseteq \vals \}$ (where $\overline{t}$ is a list of terms separated by commas) denote the set of all ground (i.e., variable-free) atoms; we refer to these as {\em facts}.
An interpretation $A\subseteq \mathcal{A}$ is a set of facts. The complete lattice $(\mathcal{P}(\mathcal{A}), \subseteq, \cap, \cup, \emptyset, \mathcal{A})$ partially orders the set of interpretations $\mathcal{P}(\mathcal{A})$.

Given a substitution $\sigma\in \vars\to \vals$, mapping variables to constants, and an atom~$a$, we write $\sigma(a)$ for the fact obtained by replacing the variables in $a$ according to $\sigma$. For example, $\sigma(p(X))$ returns the fact $p(\sigma(X))$. Given a program~$P$,
its consequence operator $T_P\in \mathcal{P}(\mathcal{A})\to \mathcal{P}(\mathcal{A})$ is defined as:
\[
T_P(A) = \{\sigma(a)\mid (a\Leftarrow l_1\ldots l_n) \in P, \forall l_i\in \overline{l}.\ A\vdash \sigma(l_i)\}
\]
where $A \vdash \sigma(a)$ if $\sigma(a)\in A$ and $A \vdash \sigma(\neg a)$ if $\sigma(a)\not\in A$.

An input for $P$ is a set of facts constructed using $P$'s extensional predicates, i.e., those that appear only in the rule bodies.
Let $P$ be a program with strata $P_1, \ldots, P_n$ and $I$ be an input for $P$. The model of $P$ for $I$, denoted by $\sem{P}_I$, is $M_n$, where $M_0 = I$ and $M_i=\bigcap \{A\in {\sf fp}\ T_{P_i}\mid M_{i-1}\subseteq A\}$ is the smallest fixed point of $T_{P_i}$ that is greater than or equal to the lower stratum's model $M_{i-1}$.

\subsection{Facts and Inference Rules}

\tool first extracts a set of base facts that hold for every instruction. These base facts constitute a Datalog input that is fed to a Datalog program to infer additional facts about the contract. 
We use the term \emph{semantic facts} to refer to the facts derived by the Datalog program. All program elements that appear in the contract, including instruction labels, variables, fields, string and integer constants, are represented as constants in the Datalog program.

\para{Base Facts}
The base facts of our inference engine describe the instructions in the contract's control-flow graph (CFG). The base facts take the form of $\instruction{instr}(L, Y, X_1, \ldots, X_n)$, where \instruction{instr} is the instruction name, $L$ is the instruction's label, $Y$ is the variable storing the instruction result (if any), and $X_1, \ldots, X_n$ are variables given to the instruction as arguments (if any).
For example, the instruction \labelc{l1: }\code{a = 4} (from Fig.~\ref{fig:overview}) is encoded to $\instruction{assign}(\text{\labelc{l1}}, \code{a}, \code{4})$. Further, the instruction 
 \labelc{l6:} \instruction{sstore}\text{(\code{c}, \code{b})}, where the variable \code{c} is known to be equal to the constant \code{0} at compile time,
is encoded to $\instruction{sstore}(\text{\labelc{l6}}, 0, \code{b})$; if the value of the variable \code{c} could not be determined at compile time, then the instruction would be encoded to $\instruction{sstore}(\text{\labelc{l6}}, \top, \code{b})$, where $\top$ is a Datalog constant that encodes that the value of \code{c} is unknown. The base facts of consecutive instructions are expressed by a predicate defined over labels called \follow. For every two labels, $L_1$ and $L_2$, whose instructions are consecutive in the CFG (either in the same basic block or in linked basic blocks), we have the base fact $\follow(L_1,L_2)$. An example, \follow~fact derived for the contract, shown in Fig~\ref{fig:overview}, is $\follow(\text{\labelc{l1}}, \text{\labelc{l2}})$.
The join of then/else branches is captured by a predicate $\relation{Join}(L_1,L_2)$, which encodes that the two branches that originate at an instruction $\instruction{goto}(L_1, X, L_3)$, located at label ~$L_1$, are joined (i.e., they are merged into a single path) at label~$L_2$.
Using the base facts described above, \tool computes two kinds of semantic facts: {\em (i)} flow-dependency predicates, which capture instruction dependencies according to the contract's CFG, and {\em (ii)} data-dependency predicates; see Fig.~\ref{fig:flow-predicates}.

\begin{figure}
\small
\renewcommand{\tabcolsep}{2pt}
\begin{tabular}{lp{165pt}}
\toprule
{\bf Semantic fact} & {\bf Intuitive meaning}\\
\midrule
\multicolumn{2}{c}{\textit{Flow Dependencies}}\\
\midrule
$\mayfollow(L_1, L_2)$ &  Instruction at label~$L_2$ may follow that at label~$L_1$.\\
$\mustfollow(L_1, L_2)$ &  Instruction at label~$L_2$ must follow that at label~$L_1$.\\
\midrule
\multicolumn{2}{c}{\textit{Data Dependencies}}\\
\midrule
$\maydependon(Y, T)$ & The value of $Y$ may depend on tag $T$.\\
$\eq(Y, T)$ & The values of $Y$ and $T$ are equal.\\
$\determinedby(Y, T)$ & For different values of $T$ the value of $Y$ is different.\\
\bottomrule
\end{tabular}
\caption{The semantic facts: $L_1$ and $L_2$ are labels, $Y$ is a variable, and $T$ is a tag (a variable or a label).}
\label{fig:flow-predicates}
\end{figure}

\para{Flow-Dependency Predicates}
The flow predicates we consider are $\mayfollow$ and $\mustfollow$, both are defined over pairs of labels and are inferred from the contract's CFG.
The intuitive meaning (also summarized in Fig.~\ref{fig:flow-predicates}) is: 
\begin{itemize}
\item $\mayfollow(L_1,L_2)$ holds for~$L_1$ and~$L_2$ if both are in the same basic block and $L_2$ follows $L_1$, or there is a path from the basic block of $L_1$ to the basic block of $L_2$.
\item $\mustfollow(L_1,L_2)$ holds if both are in the same basic block and $L_2$ follows $L_1$, or any path to the basic block of $L_2$ passes through the basic block of $L_1$.
\end{itemize}

To infer the \relation{MayFollow} and \relation{MustFollow} predicates, we use the $\follow(L_1,L_2)$ input fact which holds if $L_2$ immediately follows $L1$ in the CFG.
Namely, the predicate $\mayfollow$ is defined with the following two Datalog rules:
\[
\begin{array}{lll}
\mayfollow(L_1, L_2) &\Leftarrow& \bodyf{\follow}(L_1,L_2)\\
\mayfollow(L_1, L_2) &\Leftarrow& \bodyf{\mayfollow}(L_1,L_3),\bodyf{\follow}(L_3,L_2)\\
\end{array}
\]
The first rule is interpreted as: if $\bodyf{\follow}(L_1,L_2)$ holds (i.e., it is contained in the Datalog input), then the predicate $\mayfollow(L_1, L_2)$ is derived (i.e., it is added to the fixed-point).
The second rule is interpreted as: if both $\mayfollow(L_1, L_3)$ and $\bodyf{\follow}(L_3,L_2)$ hold, then $\mayfollow(L_1, L_2)$ is derived.
Note that if $\mayfollow(L_1, L_2)$ is not derived in the fixed-point (at the end of the fixed-point computation), then the instruction at label $L_2$ \emph{does not} appear after the instruction at label $L_1$, \emph{in any execution} of the contract.

The inference rules for $\mustfollow$ are defined similarly, with a special attention to the join points in the CFG.

\sloppy
\para{Data-Dependency Predicates}
The dependency predicates we consider are $\maydependon$, $\eq$, and $\determinedby$.
The intuitive meaning of them (also summarized in Fig.~\ref{fig:flow-predicates}) is: 
\begin{itemize}
\item $\maydependon(Y, T)$ is derived if the value of variable~$Y$ depends on the {\em tag}~$T$. Here, the variable $T$ ranges over tags, which can be a contract variable (e.g., $x$) or an instruction (e.g., \instruction{timestamp}). For example, $\maydependon(Y, X)$ means that the value of variable $Y$ may change if the value of~$X$ changes, while $\maydependon(Y, \instruction{timestamp})$ means that $Y$ may change if the instruction $\instruction{timestamp}$ returns a different value.
\item $\eq(Y, T)$ indicates that the values of $Y$ and $T$ are identical. For example, given fact $\instruction{assign}(\labelc{l}, x, \instruction{caller})$, which stores the sender's address at variable $x$, we have $\eq(x, \instruction{caller})$.
\item $\determinedby(Y, T)$ indicates that a different value of~$T$ guarantees that the value of $Y$ changes. For example, $\determinedby(x,\instruction{caller})$ is derived if we have the fact $\instruction{sha3}(\labelc{l}, \code{x}, \code{start}, \code{len})$, which returns the hash of the memory segment starting at offset \code{start} and ending at offset $\code{start} + \code{len}$, if any part of this memory segment is determined by \instruction{caller}. Note that $\eq(Y,T)$ implies that $\determinedby(Y,T)$ also holds.
\end{itemize}

\begin{figure}[tb!]
\small
\centering
\[
\arraycolsep=0pt\def\arraystretch{1.2}
\begin{array}{lll}
\toprule
\multicolumn{3}{c}{\text{\bf Data-dependency may-analysis}}\\
\midrule
\relation{MayDepOn}(Y, X)&\Leftarrow\ \ &\instruction{assign}(\_, Y, X) \\
\relation{MayDepOn}(Y, T)&\Leftarrow&\instruction{assign}(\_, Y, X), \relation{MayDepOn}(X, T)\\
\relation{MayDepOn}(Y, T)&\Leftarrow&\instruction{op}(\_, Y, \ldots, X, \ldots), \relation{MayDepOn}(X, T)\\
\relation{MayDepOn}(Y, T) &\Leftarrow&\instruction{mload}(L,Y,O),\relation{isConst}(O), \relation{MemTag}(L,O,T)\\
\relation{MayDepOn}(Y, T)&\Leftarrow&\instruction{mload}(L,Y,O),\neg\relation{isConst}(O), \relation{MemTag}(L, \_, T)\\
\relation{MayDepOn}(Y, T)&\Leftarrow&\instruction{mload}(L,Y,O), \relation{MemTag}(L,\top,T)\\
\relation{MayDepOn}(Y, T)&\Leftarrow&\instruction{assign}(L, Y, \_), \relation{Taint}(\_,L,X), \maydependon(X, T)\\

\midrule
\multicolumn{3}{c}{\text{\bf Memory analysis inference}}\\
\midrule

\relation{MemTag}(L,O,T)&\Leftarrow& \instruction{mstore}(L,O,X),\relation{isConst}(O), \relation{MayDepOn}(X, T)\\
\relation{MemTag}(L,\top,T)&\Leftarrow& \instruction{mstore}(L,O,X), \neg\relation{isConst}(O), \relation{MayDepOn}(X, T)\\
\relation{MemTag}(L,O,T)&\Leftarrow& \follow(L_1,L), \relation{MemTag}(L_1,O,T),\\
&& \neg\relation{ReassignMem}(L,O)\\
\relation{ReassignMem}(L,O)&\Leftarrow& \instruction{mstore}(L,O,\_), \relation{isConst}(O) \\

\midrule
\multicolumn{3}{c}{\text{\bf Implicit control-flow analysis}}\\
\midrule

\relation{Taint}(L_1,L_2,X)&\Leftarrow& \instruction{goto}(L_1,X,L_2)\\
\relation{Taint}(L_1,L_2,X)&\Leftarrow& \instruction{goto}(L_1,X,\_),\follow(L_1,L_2)\\
\relation{Taint}(L_1,L_2,X)&\Leftarrow& \relation{Taint}(L_1,L_3,X), \follow(L_3,L_2), \neg\relation{Join}(L_1,L_2)\\
\bottomrule
\end{array}
\]
\caption{Partial inference rules for \maydependon: the Datalog variable $X$ ranges over contract variables, $L$ ranges over instruction labels, $\top$ represents an unknown offset, and $T$ ranges over tags.}
\label{fig:inference-rules}
\end{figure}

The Datalog rules defining these data-dependency predicates are given in Fig.~\ref{fig:inference-rules}. To avoid clutter in the rules, we use the wildcard symbol (\code{_}) in place of variables that appear only once in the rule; for example, we write $\relation{MayDepOn}(Y, X)\Leftarrow \instruction{assign}(\_, Y, X)$ instead of  $\relation{MayDepOn}(Y, X)\Leftarrow \instruction{assign}(L, Y, X)$. The rules rely on additional predicates: \relation{isConst}, \relation{MemTag} (and, similarly, \relation{StorageTag}, which are omitted from Fig.~\ref{fig:inference-rules})
 and \relation{Taint}. We briefly explain the meaning of these predicates and how they are derived below.

\begin{itemize}
\item The predicate $\relation{isConst}(O)$ holds if $O$ is constant that appears in the contract. For example, the fact $\relation{isConst}(\code{0})$ is added to the Datalog input derived for the contract in Fig.~\ref{fig:overview}.
\item The predicate $\relation{MemTag}(L, O, T)$ (and similarly \relation{StorageTag}) defines that, at label $L$, the value at offset $O$ in the memory (or storage) is assigned tag $T$. It is defined with three rules. The first rule encodes that writing a variable $X$ tagged with~$T$ to a constant (i.e., known) offset~$O$ at label~$L$, results in tagging the memory offset~$O$ at label~$L$ with tag~$T$. The second rule defines the case when the offset is unknown, in which case all possible offsets, captured via the constant~$\top$, are assigned tag~$T$.
The third rule propagates the tags to the following instructions,
until reaching to an instruction that reassigns that memory location (captured by a predicate $\instruction{ReassignMem}$).
\item The predicate \relation{Taint}$(L_1,L_2,X)$ encodes that the execution of the instruction at label~$L_2$ depends on the value of~$X$, where~$X$ is the condition of a \instruction{goto} instruction at label~$L_1$.
The first two rules defining the predicate $\instruction{Taint}(L_1,L_2,X)$ taint the two branches that originate at a \instruction{goto} instruction at label~$L_1$ with the condition~$X$.
Finally, the third rule propagates the tag~$X$ along the instructions of the two branches until they are merged.
\end{itemize}

$\maydependon(X,T)$ defines that variable $X$ may have tag $T$.
The first rule defines that assigning a variable $X$ to $Y$ results in tagging $Y$ with $X$. The second rule propagates any tags of~$X$ to the assigned variable~$Y$.
The third rule propagates tags over operations with tagged variables.
The three rules with \instruction{mload} instructions propagate tags from memory to variables. The first \instruction{mload} rule defines that when loading data from a constant offset~$O$, the tags associated to that offset are propagated to the output variable~$Y$. The second \instruction{mload} rules states that if the offset is unknown, then all tags of the memory are propagated to the output variable~$Y$. Finally, the third \instruction{mload} rule propagates tags that are written to unknown offsets (identified by~$\top$).
The final rule defines that if the execution of an $\instruction{assign}(L,Y,\_)$ instruction depends on a variable~$X$ (i.e., the label~$L$ is tainted with the variable~$X$), then all tags assigned to~$X$ are propagated to the output variable~$Y$.

We remark that the rules for inferring $\eq$  and $\determinedby$ predicates are defined in a similar way and are therefore omitted.
\section{Security Patterns} \label{sec:patterns}

In this section, we show how to express security patterns over semantics facts.
We begin by defining the \tool language for expressing security patterns. 
Then, to define security properties formally, we provide background on the execution semantics of EVM contracts and formally define properties.
We continue by
presenting a set of relevant security properties, and for each, we show compliance and violation patterns, which imply the property and, respectively, its negation. This construction enables us to determine whether a contract complies with or violates a given security property.
Finally, we show how \tool leverages some patterns for error-localization.

\subsection{\tool Language}

We first define the syntax of the language for writing patterns and then define how patterns are interpreted over the semantic facts derived for a given contract (described in Section~\ref{sec:analysis}).

\para{Syntax}
The syntax of the \tool language is given by the following BNF:
\[
\begin{array}{lcl}
\varphi& ::= & \instruction{instr}(L, Y, X, \ldots, X) \mid \eq(X, T) \mid \determinedby(X, T) \\
& \mid & \maydependon(X, T) \mid \mayfollow(L, L) \mid \mustfollow(L, L)  \\
& \mid & \follow(L, L) \mid \exists X. \varphi \mid \exists L. \varphi \mid \exists T. \varphi \mid \neg \varphi \mid \varphi \wedge \varphi\\
\end{array}
\]
Here, $L$, $X$, and $T$ are variables that range over program elements such as labels, contract variables, and tags.
Patterns can refer to instructions $\instruction{instr}(L, Y, X_1, \ldots, X_n)$, where \instruction{instr} is the instruction name, $L$ is the instruction's label, $Y$ is the variable storing the instruction result (if any), and $X_1, \ldots, X_n$ are variables given to the instruction as arguments (if any).
Patterns can also refer to flow- and data-dependency semantic facts, which can be used to impose conditions on the labels and variables that appear in instructions. 
Finally, the patterns can quantify over labels, variables, and tags using the standard exists quantifier ($\exists$).
More complex patterns can be written by composing simpler patterns with negation~($\myneg$) and conjunction~($\wedge$). 

We define several syntactic shorthands that simplify the specification of patterns. We use standard logical equivalences: we write $\forall X.\ \varphi(X)$ for $\neg (\exists X.\ \neg \varphi(X))$, $\varphi_1 \vee \varphi_2$ for $\neg (\neg \varphi_1 \wedge \neg \varphi_2)$, and $\varphi_1 \Rightarrow \varphi_2$ for $\neg \varphi_1 \vee \varphi_2$. We also write $X = T$ for $\eq(X,T)$. For readability, we write:
$\myexists\ \instruction{instr}(\overline{X}).\ \varphi(\overline{Y})$
for
$\exists \overline{X}.\ \instruction{instr}(\overline{X}) \wedge \varphi(\overline{Y})$, which imposes that there is some instruction $\instruction{instr}(\overline{X})$ for which the logical condition $\varphi(\overline{Y})$ holds.
Similarly, we write $\myforall\ \instruction{instr}(\overline{X}).\ \varphi(\overline{Y})$
for
$\forall \overline{X}.\ \instruction{instr}(\overline{X}) \Rightarrow \varphi(\overline{Y})$, which imposes that for all instructions $\instruction{instr}(\overline{X})$ the condition $\varphi(\overline{Y})$ must hold.

\para{Semantics}
Patterns are interpreted by checking the inferred semantic facts:
\begin{itemize}
\item Quantifiers and connectors are interpreted as usual.
\item Flow- and data-dependency predicates are interpreted as defined in Section~\ref{sec:analysis}; i.e., a semantic fact holds if and only if it is contained in the Datalog fixed-point.
\end{itemize}
For example, consider the pattern:
\[
\myexists \sstorepc{X}{Y}{L}.\ \determinedby(X, \instruction{caller})
\]
which is a shorthand for $\exists X.\ \sstorepc{X}{Y}{L} \wedge \determinedby(X, \instruction{caller})$.
This pattern is matched if there is an instruction $\instruction{sstore}(L,X,Y)$ in the contract such that the offset $X$ is determined by the address returned by the \instruction{caller} instruction (captured by the predicate $\determinedby(X, \instruction{caller})$).
For brevity, we omit variables that are not conditioned in the pattern:
$\myexists \sstorepc{X}{\_}{\_}.\ \determinedby(X, \instruction{caller})$.

In Fig.~\ref{fig:patterns}, we list security patterns that are built-in in \tool. In the following, we first give additional background on the EVM execution model and then present these patterns.

\subsection{EVM Background and Properties}
To understand the security properties defined in the next section, we extend the background on EVM (given in Section~\ref{sec:background}), which focused on the EVM syntax, with the semantics 
of EVM contracts.

\para{EVM Semantics}
A contract is a sequence of EVM instructions $C=(c_0,\ldots,c_m)$.
The semantics of a contract $\sem{C}$ is the set of all \emph{traces} from an initial state. A trace of a contract $C$ is a sequence of state-instruction pairs $(\sigma_0,c_0) \to \ldots \to (\sigma_k,c_k)$, from an initial state $\sigma_0$, and such that the relation $(\sigma_j,c_j) \to (\sigma_{j+1},c_{j+1})$ is valid according to the EVM execution semantics~\cite{wood2014ethereum}. 
If a trace successfully terminates, then $c_k=\bot$.
A state consists of the storage and memory state (mentioned in Section~\ref{sec:background}), stack state, transaction information, and block information. We denote by $\sigma_{\storage[i]}/\sigma_{\heap[i]}$ the value stored at offset $i$ in the storage/memory,
by $\sigma_{\storage}/\sigma_{\heap}$ the state of the storage/memory,
 by $\sigma_{\balances}$ the contract's balance,
by $\sigma_{T}$ the transaction, and by $\sigma_{B}$ the block information.
We denote by $t[i]$ the $i^\text{th}$ pair of the trace $t$, for a positive $i$. For a negative $i$, $t[i]$ refers to the $i^\text{th}-1$ pair of $t$ from the end of the sequence.
We denote by $\sigma^{t[i]}$/$c^{t[i]}$ the state/instruction of the $i^\text{th}$ pair of $t$, and by $\sigma^{t[i]}_f$ the value of instruction $f$ (e.g., \instruction{caller}) in $\sigma^{t[i]}$. 

\para{Properties}
A property is a relation over sets of traces.
A contract satisfies a security property $\rho$ if $\sem{C}\in \rho$. If $\sem{C}\notin \rho$, we say that $C$ violates the property $\rho$.
We define relations using first-order logic formulas. The formulas are interpreted over the traces and the bitstrings that comprise the user identifiers, offsets, and other arguments or return values of the EVM instructions. We denote by $t_1,t_2,...$ variables that refer to traces. We denote by $i_1,i_2,...$ variables that refer to the index of a pair in a trace. We use other letters for bitstring variables. For example, we use $a$ to refer to a bitstring which is used in the formula to refer to a user's identifier (her address), and we use $x$ to refer to an offset in the storage or as arguments to \instruction{call}.
For simplicity's sake, although EVM is a stack-based language, we write instructions as $r\leftarrow \instruction{instr}(a_1,\ldots,a_k)$ and use the wildcard for arguments/return values that are not important to the formula. Note that $a_1,...a_k,r$ represent the concrete values at the moment of execution.

\begin{figure*}[htb!]
\centering
\def\arraystretch{1.1}%
\begin{tabular}{lll}
\toprule
{\bf Property} & {\bf Type} & {\bf Security Pattern} \\
\midrule
\textbf{LQ: Ether}& {\em compliance} &
$\myforall \sstoppc{L_1}.\ \myexists \gotopc{X}{L_3}{L_2}.\ X = \val \myand \follow(L_2, L_4) \myand L_3 \neq L_4 \myand \mustfollow(L_4, L_1)$\\
\textbf{liquidity}&{\em compliance} & $\myexists \callpc{\_}{\_, Amount}{L_1}. Amount \neq 0 \myor \determinedby(Amount, \instruction{data})$\\ 
& {\em violation} & $\big(\myexists \sstoppc{L}.\  \myneg \maydependon(L, \instruction{callvalue})\big)\ \myand\ \big(\myforall \callpc{\_}{\_, Amount}{\_}.\  Amount = 0\big)$ \\

\midrule
{\bf NW: No writes} & {\em compliance}  & $ \myforall \callpc{\_}{\_,\_}{L_1}.\ \myforall \sstorepc{\_}{\_}{L_2}.\  \neg \mayfollow(L_1, L_2)$\\
{\bf after call}& {\em violation} & $\myexists \callpc{\_}{\_,\_}{L_1}.\ \myexists \sstorepc{\_}{\_}{L_2}.\ \mustfollow (L_1, L_2)$\\
\midrule
\textbf{RW: Restricted} & {\em compliance} & $\myforall \sstorepc{X}{\_}{\_}.\  \determinedby(X, \instruction{caller})$ \\
\textbf{write}& {\em violation} & $\myexists \sstorepc{X}{\_}{L_1}.\  \myneg \maydependon(X, \instruction{caller}) \myand \myneg \maydependon(L_1, \instruction{caller})$\\
\midrule
\textbf{RT: Restricted} & {\em compliance} & $\myforall \callpc{\_}{\_, Amount}{\_}.\  Amount = 0$\\
\textbf{transfer} 
& {\em violation} & $\myexists \callpc{\_}{\_, Amount}{L_1}.\  \determinedby(Amount, \instruction{data}) \myand \myneg \maydependon(L_1, \instruction{caller})\myand \myneg  \maydependon(L_1, \instruction{data})$\\
\midrule
{\bf HE: Handled} & {\em compliance} & $\myforall \callpc{Y}{\_,\_}{L_1}.\  \myexists \gotopc{X}{\_}{L_2}.\  \mustfollow(L_1, L_2)  \myand \determinedby(X, Y)$\\
\textbf{exception} & {\em violation} & $\myexists \callpc{Y}{\_,\_}{L_1}.\  \myforall \gotopc{X}{\_}{L_2} .\  \mayfollow(L_1, L_2) \myimplies \myneg \maydependon(X, Y)$\\
\midrule
{\bf TOD: Transaction} & {\em compliance} & $\myforall \callpc{\_}{\_, Amount}{\_}.\  \myneg \maydependon(Amount, \instruction{sload}) \myand  \myneg \maydependon(Amount, \instruction{balance}) $\\
{\bf ordering} & {\em violation} & $\myexists \callpc{\_}{\_,Amount}{\_}.\  \myexists \sloadpc{Y, X}{\_}.\  \myexists \sstorepc{X}{\_}{\_}.\ \determinedby(Amount, Y) \myand \relation{isConst}(X) $\\
{\bf dependency}
 &  & \\
\midrule

{\bf VA: Validated} & {\em compliance} & $ \myforall \instruction{sstore}(L_1, \_, X).\  \maydependon(X, \instruction{arg})$\\
{\bf arguments} && $\qquad \myimplies \big(\myexists \instruction{goto}(L_2, Y, \_).\ \mustfollow(L_2, L_1) \myand \determinedby(Y, \instruction{arg})\big)$\\

 & {\em violation} & $\myexists \instruction{sstore}(L_1, \_, X).\ \determinedby(X, \instruction{arg}) $\\
&&$\qquad \myimplies \neg \big(\myexists \instruction{goto}(L_2, Y, \_).\ \mayfollow(L_2, L_1) \myand \maydependon(Y, \instruction{arg})\big)  $\\

\bottomrule
\end{tabular}
\caption{Compliance and violation security patterns for relevant security properties}
\label{fig:patterns}
\end{figure*}

\subsection{Security Properties and Patterns}\label{sec:patts}

We now define seven security properties with respect to the EVM semantics~\cite{wood2014ethereum}. 
Checking these properties precisely is impossible since EVM is Turing-complete. Instead, for each property, we define compliance and violation patterns over our language, which over-approximate the property and, respectively, its negation. That is, a compliance pattern match implies that the property holds, and a violation pattern match implies that the property's negation holds. If neither pattern is matched, then the property may or may not hold. 
In the following, for each security property, we describe its relevance, present its formal definition, and then refine it into a set of compliance and violation patterns. The complete list of properties and patterns is given in Fig.~\ref{fig:patterns}.

\para{Ether Liquidity (LQ)} In November 2017, a bug in a contract led to freezing $\$160$M~\cite{parity2}. The bug occurred because a contract relied on another smart contract (acting as a library) to transfer its ether to users. Unfortunately, a user accidentally removed the library contract, freezing the contract's ether. The combination of the contract being able to receive ether from users and the absence of an explicit transfer to the user led to this issue.
Formally, we define this security property by requiring that {\em (i)} all traces $t$ do not change the contract's balance (which means that the contract has no ether and thus its ether is vacuously liquid), or {\em (ii)} there exists a trace $t$ that decreases the contract's balance (i.e., ether is liquid).
\[
\begin{array}{l}	
\psi_{LQ}=(\forall t.
\sigma_{\balances}^{t[0]} = \sigma_\textit{\balances}^{t[-1]})
~\vee
(\exists t.
\sigma_{\balances}^{t[0]} > \sigma_\textit{\balances}^{t[-1]})
\end{array}
\]
To over-approximate $\psi_{LQ}$ with our language, we leverage the fact that if ether is transferred to the contract, then the amount of ether transferred is given by the $\val$ instruction. Thus, if, for all traces that complete successfully, this amount is zero, then the first part of $\psi_{LQ}$ is satisfied.
These is exactly the first liquidity compliance pattern in Fig.~\ref{fig:patterns}: it matches if all transactions that can complete successfully (reach a \instruction{stop} instruction) have to follow a branch of a condition (where the condition is identified by a \instruction{goto} instruction) that is reachable only if the ether transferred to this contract is zero (this branch is the one to which the \instruction{goto} instruction does not jump).
The second liquidity compliance pattern over-approximates the second part of $\psi_{LQ}$. It leverages the fact that ether is liquid if there is a reachable \instruction{call} instruction which sends a non-zero amount of ether. Concretely, it is matched if there is a \instruction{call} instruction which transfers  {\em (i)}~a positive amount of ether or {\em (ii)}~amount of ether  which depends only on the transaction data, and thus can be positive. 

Our violation pattern over-approximates $\neg\psi_{LQ}$ by checking that both conditions are false: the contract can receive ether, but cannot transfer ether.
To guarantee that the contract can receive ether, it verifies that there is an execution that can complete successfully (i.e., reach \instruction{stop}) and its execution does not depend on \instruction{callvalue} -- this guarantees that some trace with positive \val\ can complete. To guarantee that ether cannot be transferred, it verifies that all \instruction{call} instructions transfer $0$ ether.

\para{No Writes After Calls (NW)} In July 2016, a bug in the DAO contract enabled an attacker to steal $\$60$M~\cite{daocite}. The attacker exploited the combination of two factors. First, a \instruction{call} instruction which upon its execution enabled the recipient of that \instruction{call}
to execute her own code before returning to the contract. Second, the amount transferred by this \instruction{call} depended on a storage value, which was updated \emph{after} this \instruction{call}. This value was critical as it recorded the 
amount of ether that the
\instruction{call}'s recipient had in the contract, and can thus request to receive. 
This allowed the attacker to call the function again \emph{before the storage was updated}, thus making the contract believe that the user still had ether in the contract.
A property that captures when this attack cannot occur checks that there are no writes to the storage after any $\instruction{call}$ instruction. 
We formalize this vulnerability by requiring that, for all traces $t$,
the storage does not change in the interval that starts just before any \instruction{call} instruction and ends when the trace completes:
\[
\psi_{NW}=\forall t \forall i  (i<-1 \wedge c^{t[i]} = \_ \gets \instruction{call}(\_, \_, \_)) 
\Rightarrow\sigma_{\storage}^{t[i]} = \sigma_{\storage}^{t[-1]}
\]
Note that this property is different from reentrancy~\cite{luu2016making}, which stipulates that the callee must not be able to re-enter the same function and reach the \instruction{call} instruction.
Our compliance rule over-approximates $\psi_{NW}$ by leveraging the fact that the storage can only be changed via $\instruction{sstore}$. It is thus matched if \instruction{call} instructions are not followed by $\instruction{sstore}$ instructions. 
Our violation pattern over-approximates $\neg \psi_{NW}$ by checking that there is a \instruction{call} instruction which must be followed by a write to the storage, in which case the implication of $\psi_{NW}$ is violated. 

\para{Restricted Writes (RW)}
In July 2017, an attacker stole $\$30$M because of an unrestricted write to the storage~\cite{paritybug}.
The attacker exploited the reliance of the contract on a library that enabled to unconditionally set an \emph{owner} field to any address. This enabled the attacker to take ownership over the contract and steal its ether.
We consider a security property that guarantees that writes to storage are restricted.
The property requires that, for every storage offset $x$ (e.g., a field in the contract), there is a user $a$ that cannot write at offset $x$ of the storage. 
\[
\begin{array}{l}	
\psi_{RW}=\forall x  \exists a \forall {t}
(\sigma_{\instruction{caller}}^{t[0]} = a  \Rightarrow c^{t[-1]} \neq \instruction{sstore}(\_, x,\_))\\

\end{array}
\]
Our compliance pattern over-approximates $\psi_{RW}$ by checking that offsets of \instruction{sstore} instructions, denoted $x$, are determined by the sender's identifier (i.e., users can only write to their designated slot). This ensures that for all $x$, there exists a user $a$ (in fact, all users but one) who cannot write to $x$.
The violation pattern over-approximates $\neg \varphi_{RW}$ by checking if there is an \instruction{sstore} instruction whose execution and offset are independent of \instruction{caller}. In this case, we can define an offset $x$, for which all users can write -- hence violating the property.
If this property is too restrictive (there are cases where it is safe to allow global writes to the storage), one can define it (and adapt the patterns) with respect to critical writes (e.g., writes that modify an \code{owner} field), identified by their label $l$.

In the following, we skip the formal definition of properties, and only describe them informally.

\para{Restricted Transfer (RT)}
We define a property that guarantees that ether transfers (via \instruction{call}) cannot be invoked by any user $a$. Violation of this property can detect Ponzi schemes~\cite{BartolettiCCS17}.
Our compliance pattern requires that for all users, invocations of that \instruction{call} instruction do not transfer ether.
Our first violation pattern checks if the \instruction{call} instruction transfers non-zero amount of ether and its execution is independent of the sender. For the second violation pattern, the amount of ether transferred depends on the transaction data (and thus can be set to a non-zero value), while the execution is independent of this data (and will thus take place).

\para{Handled Exception (HE)} In February 2016, a contract by the name ``King of Ether'' had an issue due to mishandled exceptions, forcing its creator to publicly ask users not to send ether to it \cite{koe2}. 
The issue was that the return value of a \instruction{call}, which indicated if the instruction completed successfully, was not checked. 

Our compliance pattern checks that \instruction{call} instructions are followed by a \instruction{goto} instruction whose condition is determined by the return code of \instruction{call}. This guarantees that depending on the return code, different execution paths are taken.
Our violation pattern checks that the \instruction{call} instruction is not followed by a \instruction{goto} instruction which may depend on the return value. This guarantees that there is no different behavior depending on the result of the \instruction{call}.

\para{Transaction Ordering Dependency (TOD)}
An inherent issue in the blockchain model is that there is no guarantee on the execution order of transactions.
While this has been known, it recently became critical in the context of Initial Coin Offerings, a popular means for start-ups to collect money by selling tokens.
The initial tokens are sold at a low price while offering a high bonus, and as demand increases the price increases and the bonus decreases.
It has been observed that miners exploit this to
create their transactions to win the big bonus at a low rate~\cite{ico}.

Our compliance pattern requires that the amount of ether send by a \instruction{call} instruction is independent of the state of the storage and contract's balance. This means that reordering transactions (which can be affected by changing the storage or balance) does not affect the amount sent by the \instruction{call} execution. Our violation pattern checks that the amount of the \instruction{call} instruction is determined by a value read from the storage, whose offset in the storage is known (i.e., it is constant), and that this value can be updated.

In Section~\ref{sec:evaluation}, we evaluate several versions of the TOD property:
\begin{inparaenum}[\em (i)]
\item{\em TOD Transfer (TT)} indicates that the execution of the ether transfer depends on transaction ordering (e.g., a condition guarding the transfer depends on the transaction ordering);
\item{\em TOD Amount (TA)} marks that the amount of ether transferred depends on the transaction ordering (this variation is the one described above and in Fig.~\ref{fig:patterns});
\item{\em TOD Receiver (TR)} captures the vulnerability that the recipient of the ether transfer might change, depending on the transaction ordering.
\end{inparaenum}

\para{Validated Arguments (VA)} Method arguments should be validated before usage, because unexpected arguments may result in insecure contract behaviors. Contracts must check whether all transaction arguments meet their desired preconditions.

Our compliance pattern checks that before storing in the persistent memory a variable that may depend on a method argument, there exists a check of the argument value. Our violation pattern identifies \instruction{sstore} instructions that write to memory a 
method argument without previously checking its value.

\para{Limitations}
We next discuss a few limitations of checking properties through patterns.
First, all our violation patterns assume that the violating instructions (which match the violation pattern) are part of some terminating execution.
For example, in the violation pattern of ether liquidity, the matching \instruction{stop} is assumed to be reachable, and in the violation pattern of no writes after calls, both the \instruction{call} and the write are assumed to be part of some terminating execution.
We take this assumption since, in general, this problem is undecidable.

Second, the security properties we consider are generic and do not capture contract-specific requirements (we illustrate the specification of contract-specific patterns in \tool's DSL below). Some vulnerabilities are, however, contract-specific, and therefore they are not captured by our compliance patterns (i.e., a contract can be exploitable even if a compliance pattern is matched).
For example, our compliance pattern for handled exceptions matches if there is \emph{some} check over the \instruction{call}'s return value. However, the pattern cannot check that the exception was handled \emph{correctly}, as this is contract-specific. 
Similarly, the compliance pattern for validated arguments matches if there is \emph{some} check over the arguments. However, the check can still miss cases where inputs are not correctly validated, as the meaning of \emph{correctly validated} varies across contracts.

Third, since our patterns do not capture precisely their corresponding properties, it can happen that a contract matches neither the compliance nor the violation pattern. 
In this case, \tool cannot infer whether the property holds, and thus shows a warning.

\para{Contract-specific Patterns} 
Finally, we remark that \tool is not limited to checking the security properties described above. In fact, it is common that a security auditor would write custom patterns defined for a particular contract. Such custom patterns are specified by providing an expression in the \tool language. 

To illustrate this, suppose an auditor wants to check whether the execution of a specific sensitive \instruction{call} instruction at label~$\labelc{l}$ depends on the address of the owner. To discover violations of this property, the auditor would write: 
$$
\begin{array}{lll}
\myexists \instruction{call}(L, \_, \_, \_).\ \\
\qquad (L = \labelc{l}) \wedge \myneg \big(\myexists \instruction{sload}(\_, Owner, X).\ \maydependon(L, X)\big)
\end{array}
$$
Here, $Owner$ is the identifier of the field storing the owner address, i.e. a constant offset in the contract's storage.

\subsection{Error Localization via Violation Patterns}
An important part of \tool is to pinpoint the instructions that lead to violations (or potential violations) of security properties, as this enables developers to fix the code.
In this section, we characterize which patterns enable such error localization.
We call such patterns \emph{instruction patterns} (as they pinpoint instructions), and we call other patterns \emph{contract pattern} (as the violation is identified for the entire contract).

\para{Instruction Patterns}
An instruction pattern has the form of:
$\myexists \instruction{instr}(\overline{X}).\ \varphi_v(\overline{X})$, for violation patterns,
and, 
$\myforall \instruction{instr}(\overline{X}).\ \varphi_c(\overline{X})$, for compliance patterns.
That is, if a violation pattern is an instruction pattern and it is matched by some  $\instruction{instr}(\overline{X})$, then \tool can highlight this instruction as a violation.
Similarly, if a compliance pattern is an instruction pattern and it is \emph{not} matched because of some $\instruction{instr}(\overline{X})$, then \tool can highlight this instruction as a warning (assuming that the corresponding violation pattern has not matched).
Note that six of the violation patterns in Fig.~\ref{fig:patterns} (all except the violation pattern for ether liquidity) are instruction patterns. 

\para{Contract Patterns}
Patterns which are not instruction patterns are called \emph{contract patterns}. For them, 
it is difficult to pinpoint a single instruction responsible for its violation. The ether liquidity violation pattern is an example of a contract pattern: it conjoins two different conditions pertaining to \instruction{stop} and \instruction{call} instructions. For contract patterns, \tool evaluates the compliance and violation patterns and flags the contract as vulnerable (if the violation pattern is matched) or issues a warning (if no pattern is match) without pinpointing specific instructions.

\section{Implementation}\label{sec:implementation}

In this section, we detail the implementation of \tool.


\para{Decompiler} The decompiler transforms the EVM bytecode provided as input into the corresponding assembly instructions, as defined in~\cite{wood2014ethereum}.
Next, it converts the EVM instructions into an SSA form. The SSA instructions are identical to the EVM instruction set except that they exclude stack operations (e.g., \instruction{pop}, \instruction{push}, etc.). Our conversion method is similar to the one described in~\cite{Proebsting97krakatoa:decompilation,Vallee-Rai98jimple:simplifying}.
The decompiler constructs the control flow graph (CFG) on top of the decompiled instructions.

\para{Optimizations}
\tool employs three optimizations over the CFG, which improve the precision of its analysis:
\begin{enumerate}
\item[\em (i)]
{\em Unused instructions}, which eliminates any instructions whose results are not used.
On average, this optimization reduces the contract's instructions by $44$\% and improves the scalability and precision of the subsequent analysis.
\item[\em (ii)]
{\em Partial evaluation}, which propagates constant values along computations~\cite{Futamura:1999:PEC:609149.609205}.
This step improves the precision of storage and memory analysis (e.g., \relation{MemTag}). As we show in our evaluation, partial evaluation resolves over $70\%$ of the offsets that appear in storage/memory instructions.
 \item[\em (iii)]
{\em Method inlining}, which improves the precision of the static analysis by making it context sensitive.
\end{enumerate}

\para{Inference of Semantic Facts} \tool derives semantic facts using inference rules specified in stratified Datalog, using the Souffle Datalog solver~\cite{souffle} to efficiently compute a fixed-point of all facts. We report on concrete numbers in Section~\ref{sec:evaluation}.

\para{Evaluating Patterns}
To check the security patterns, \tool iterates over the instructions to handle the \emph{\myforall} and \emph{\myexists} quantifiers in the patterns. Then, to check inferred facts, it directly queries the fixed-point computed by the Datalog solver.
If a violation pattern is matched, \tool reports which instructions are identified as vulnerable, to provide error-localization for users. If no pattern is matched, \tool reports a warning, to indicate that an instruction may or may not be vulnerable.

\section{Evaluation}\label{sec:evaluation}

\begin{figure}
\setlength{\tabcolsep}{8pt}
\begin{tabular}{lrr}
\toprule
& {\bf EVM dataset} & {\bf Solidity dataset}\\
\midrule
\# Contracts & $24,594$ & $100$\\
\# \instruction{call} instructions & $46,106$ & $67$\\
\# \instruction{sstore} instructions & $56,346$ & $297$\\
\bottomrule
\end{tabular}
\caption{Statistics of the two Ethereum datasets}
\vspace{-10pt}
\label{fig:datasets}
\end{figure}

To evaluate \tool, we conducted the following experiments:
{\em (i)}~evaluated \tool's effectiveness in proving the correctness of and discovering violations in real-world contracts;
{\em (ii)}~ manually inspected \tool's results (i.e., reported violations and warnings) on smart contracts whose source code had been uploaded to \tool's public interface;
{\em (iii)}~ compared \tool to Oyente~\cite{luu2016making} and Mythril~\cite{mythril},  two smart contract checkers based on symbolic execution;
{\em (iv)}~ measured the success of \tool's decompiler in resolving memory and storage offsets;
{\em (v)}~ measured \tool's time and memory consumption.

\para{Datasets} \label{sec:dataset}
We used two datasets of smart contracts to evaluate \tool. Our first dataset, dubbed {\em EVM dataset}, consists of $24,594$ smart contracts obtained by parsing create transactions using the parity client~\cite{parityclient}. Using create transactions, we obtained the EVM bytecode of these smart contracts. Our second dataset, dubbed {\em Solidity dataset}, consists of $100$ smart contracts written in Solidity which were uploaded to \tool's public interface. To avoid bias, we selected the first $100$ contracts in alphabetical order uploaded in $2018$. To simplify manual inspection, we restricted our selection to contracts with up to $200$ lines of Solidity code.

We give relevant statistics on the two datasets in Fig.~\ref{fig:datasets}. 
Note that the number of contracts defines the relevant checks for the ether liquidity (LQ) property,
the number of \instruction{sstore} instructions defines the relevant instructions for the restricted writes (RW) and the validated arguments (VA) property,
and the number of \instruction{call} instructions defines the relevant instructions for the remaining properties.




\begin{figure}
\includegraphics[width=1\columnwidth]{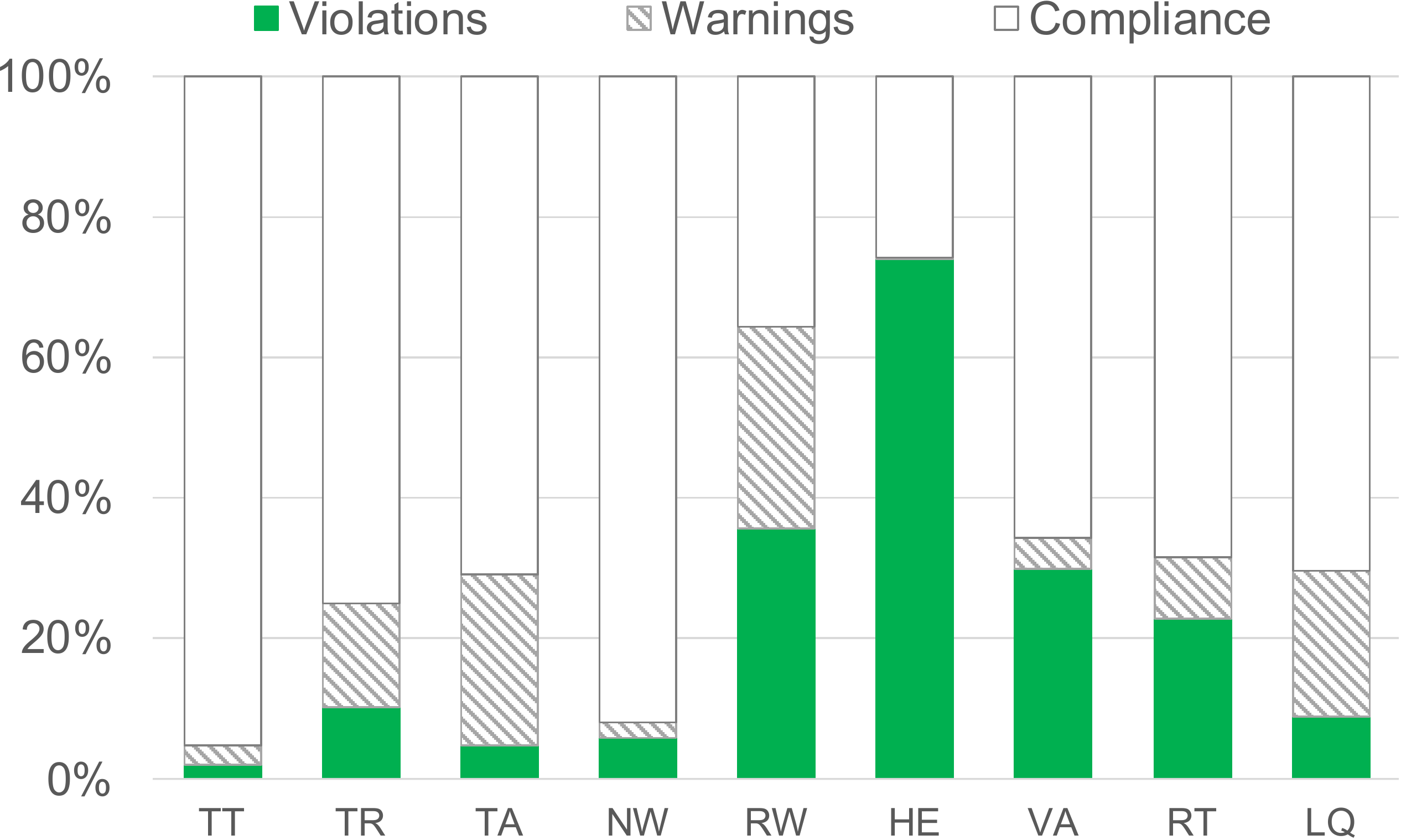}
\caption{\tool results on the EVM dataset. The violations and compliance segments indicate instructions that are proved to be safe/violations for each security property.}
\label{fig:all-contracts}
\end{figure}

\para{Security Analysis of Real-World Smart Contracts}
In this task, we evaluate \tool's effectiveness in proving security properties (i.e., matching a compliance pattern) and finding violations (i.e., matching a violation pattern) in real-world contracts. To this end, we ran \tool on all smart contracts contained in our EVM dataset and measured the fraction of violations, warnings, and compliances reported by \tool.

Fig.~\ref{fig:all-contracts} summarizes the results. The figure shows one bar for each security property. Each bar has three segments: {\em (i) violations}, which shows the fraction of instructions that have matched a violation pattern of the given property, {\em (ii) warnings}, which shows the fraction of instructions that have not matched any pattern (neither violation or compliance pattern) of the given property, and {\em (iii) compliance}, which shows the fraction of instructions that have matched a compliance pattern of the given property. We note that the sum of the three segments adds up to~$100\%$.

For example, consider the no writes after calls (NW) property. The data shows that $6.5\%$ of the \instruction{call} instructions violate the property,
$90.9\%$ are proved to be compliant, and the remaining $2.6\%$ are reported as warnings.
On average across all security properties, \tool successfully proves that $55.5\%$ of the relevant instructions are safe, $29.3\%$ are definite violations, and it reports $15.2\%$ warnings. Further, $65.9\%$ of all instructions that failed to match a compliance pattern (and hence may indicate an error) are successfully proved to be definite violations (using the violation patterns). This indicates a reduction of $65.9\%$ in the number of instructions that users must manually classify into true warnings and false warnings. We report on the precise breakdown between false and true warnings in our next experiment.

Overall, our results indicate that \tool's compliance and violations patterns are expressive enough to prove and, respectively, disprove relevant security properties. Further, we note that since \tool is extensible, one can further refine \tool's results by extending it with additional patterns that would convert more warnings into violations and compliances. This would benefit some of the security properties that are harder to prove or disprove (such as restricted writes).


\begin{figure}
\includegraphics[width=1\columnwidth]{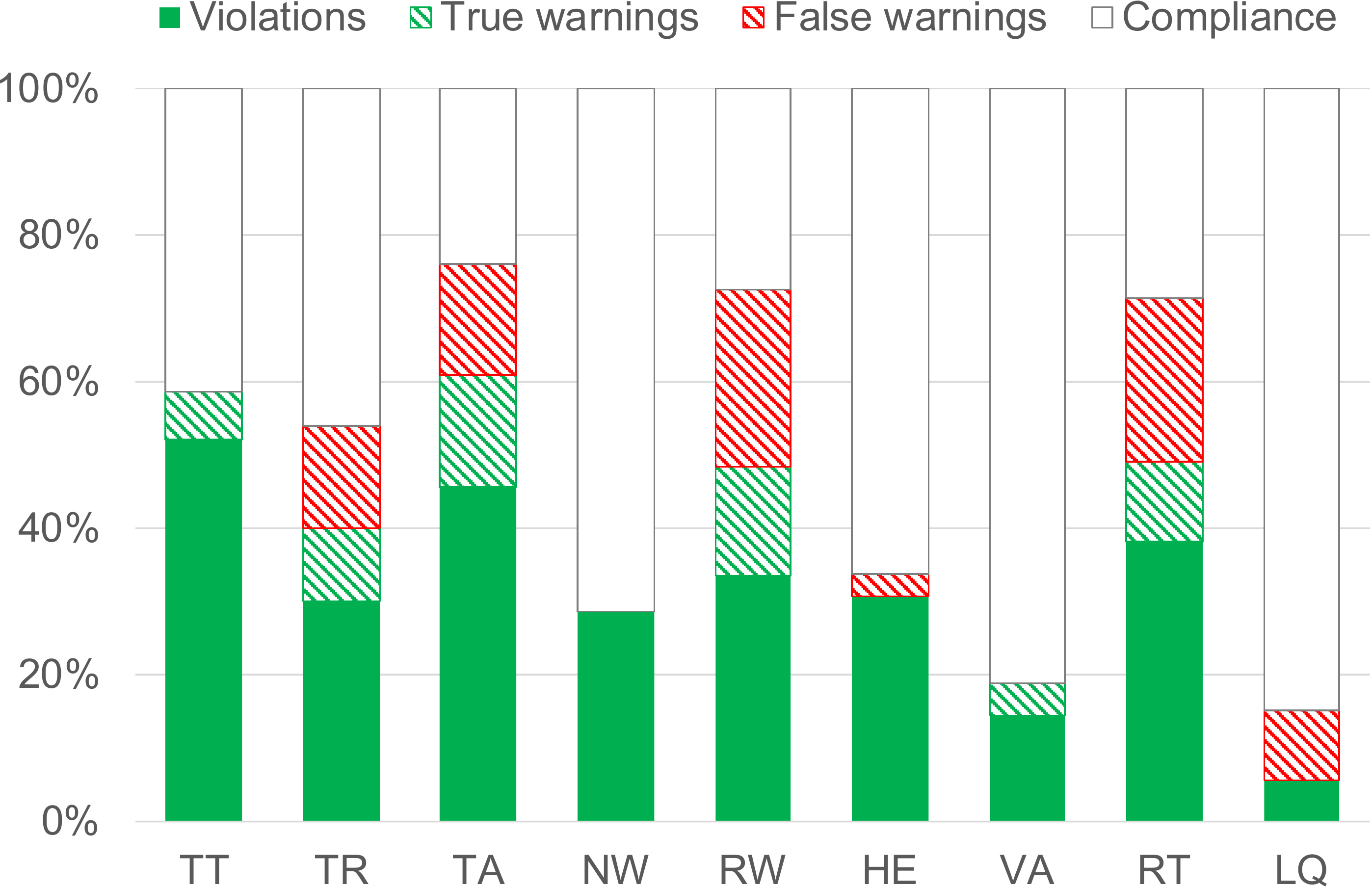}
\caption{\tool results on the Solidity dataset. The warnings are classified into true and false warnings based on whether they indicate a security issue or not.}
\label{fig:accuracy}
\end{figure}

\para{Manual Inspection of Results}
In our second experiment, we manually inspected \tool's reports to gain a better understanding of its results.
To this end, we ran \tool on all contracts contained in our Solidity dataset. We then manually classified each reported warning as a {\em true warning} if it indicates a violation of the security property, and otherwise, we classified it as a {\em false warning}. We also inspected and confirmed the correctness of all reported violations and compliances.

Fig.~\ref{fig:accuracy} summarizes our results. As before, the figure shows one bar for each security property. In addition to the violation and compliance segments, we partition the segment with reported warnings into {\em true warnings} and {\em false warnings}. 

Consider the handled exception (HE) property. The data shows that \tool successfully proves that $29.9\%$ of the \instruction{call} instructions have return values that are {\em not} checked by the code (indicating a violation of the property).  Further, \tool proves that these error values {\em are} checked for the remaining $70.1\%$ of \instruction{call} instructions. \tool does not issue any warnings for this property because it matched at least on of the patterns for each of the \instruction{call} instructions. 

We remark that the number of security issues discovered in the Solidity dataset is higher relative to those found in the EVM dataset. We believe this is due to the fact that
the two datasets come from different distributions: the Solidity dataset consists of recent contracts (uploaded in $2018$) that are still in development stage. In contrast, the EVM dataset contains all contracts deployed on the blockchain. Further, users often deliberately uploaded vulnerable contracts to experiment and evaluate \tool. An exception is the reduction in handled exception property (HE), which has more violations in the EVM dataset compared to the Solidity dataset. We believe this is due to the fact that developers now use the \code{transfer()} statement for ether transfers, which handles errors by default and was specifically introduced to avoid issues due unhandled exceptions.

We observe that the effectiveness of the patterns varies across properties, which is expected as some properties are more difficult to prove/disprove than others. For example, the restricted transfer property (RT) and the three transaction ordering dependence properties (TT, TR, and TA) are hard to prove correct and result in a relatively high number of false warnings (roughly half of the warnings are false warnings). However, for other security properties, such as no writes after calls (NW) and handled exception (HE), all warnings issued by \tool indicate true warnings, indicating that the corresponding compliance patterns precisely matches contracts that satisfy these properties.


\begin{figure}
\includegraphics[width=1\columnwidth]{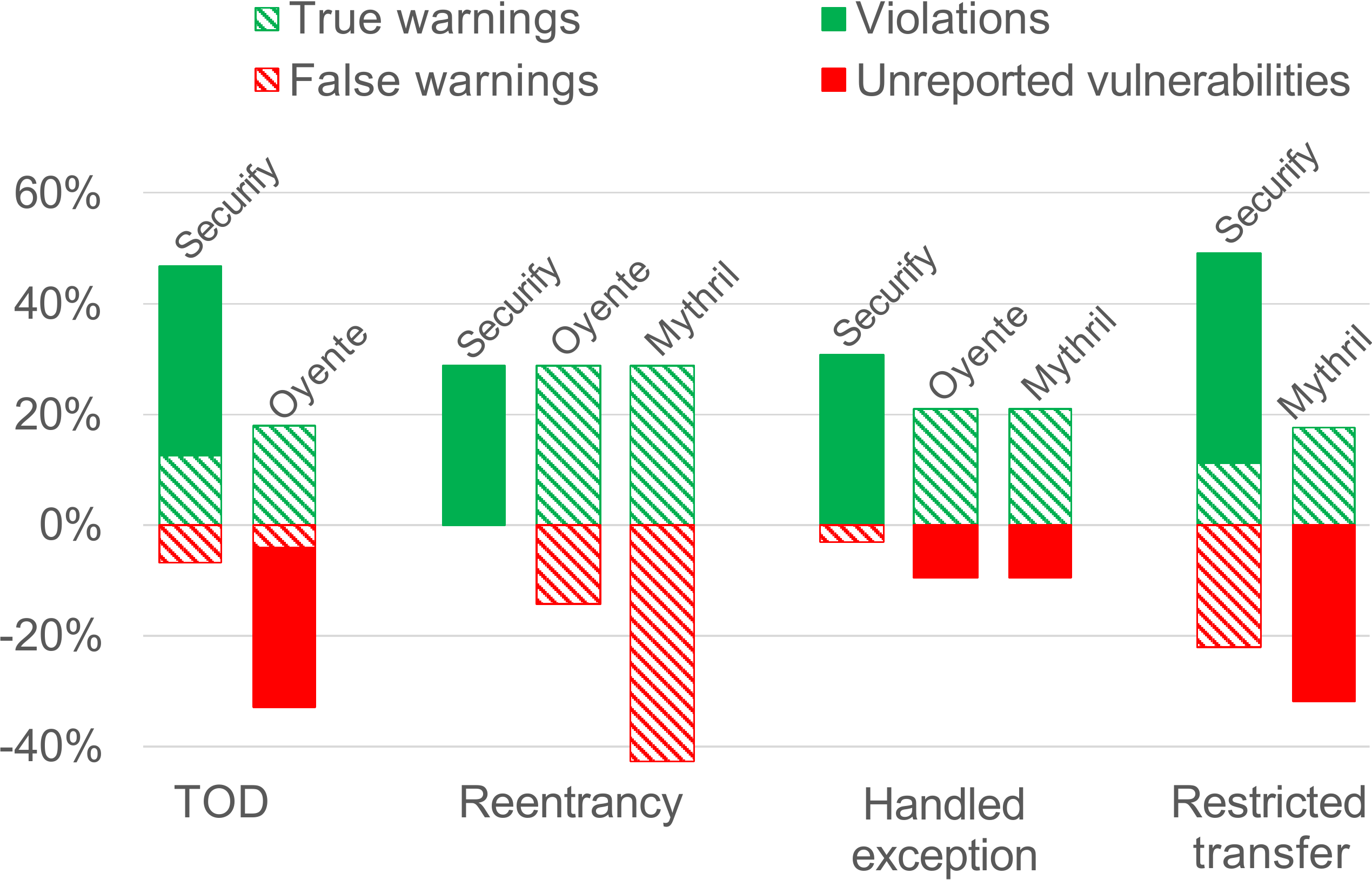}
\caption{Comparing \tool to Oyente and Mythril}
\label{fig:compare}
\end{figure}

\para{Comparing \tool to Symbolic Security Checkers}
We now compare \tool to two recent open-source security checkers based on symbolic execution -- Oyente~\cite{luu2016making} and Mythril~\cite{mythril}. 
To compare the three systems, we ran the latest versions of Oyente and Mythril against all contracts in our Solidity dataset, for which we have already manually classified all warnings into true and false warnings. Oyente supports three of \tool's security properties: TOD, which checks the disjunction of the TOD receiver and TOD amount properties, reentrancy (called no writes after calls\footnote{We remark that to ensure the absence of storage writes after \instruction{call} instructions, Oyente checks that the user cannot re-enter and reach the same \instruction{call} instruction.} in \tool), and handled exceptions. Mythril also supports the reentrancy and handled exception properties, and in addition, implements a check of the restricted transfer property.

Our results are summarized in Fig.~\ref{fig:compare}.
For \tool, we report the fraction of reported violations, true warnings, and false warnings. Since both Oyente and Mythril may report false positives (Oyente has false positives because their checks do not imply a contract vulnerability, as shown in~\cite{GrishchenkoMS18}), we treat all bugs listed by them as {\em warnings} as they must be classified by the user into true warnings and false warnings. Note that, unlike \tool, Oyente and Mythril do not report definite violations, i.e., results that are guaranteed to violate security properties. Since Oyente and Mythril explore a subset of all contract's behaviors, they may fail to report certain vulnerabilities, and we report these as {\em unreported vulnerabilities} in the figure. We depict true warnings and violations above the $X$-axis (to indicate desirable results), and we plot false warnings and unreported vulnerabilities below the $X$-axis (to indicate undesirable results).

We observe that for all properties except reentrancy, Oyente and Mythril miss to report some actual vulnerabilities. Oyente fails to report $72.9\%$ of TOD violations, and Mythril fails to report $65.6\%$ of the restricted transfer violations. 
Overall, the two symbolic tools fail to report vulnerabilities for all considered security properties.


\begin{figure}
\centering
\includegraphics[width=0.8\columnwidth]{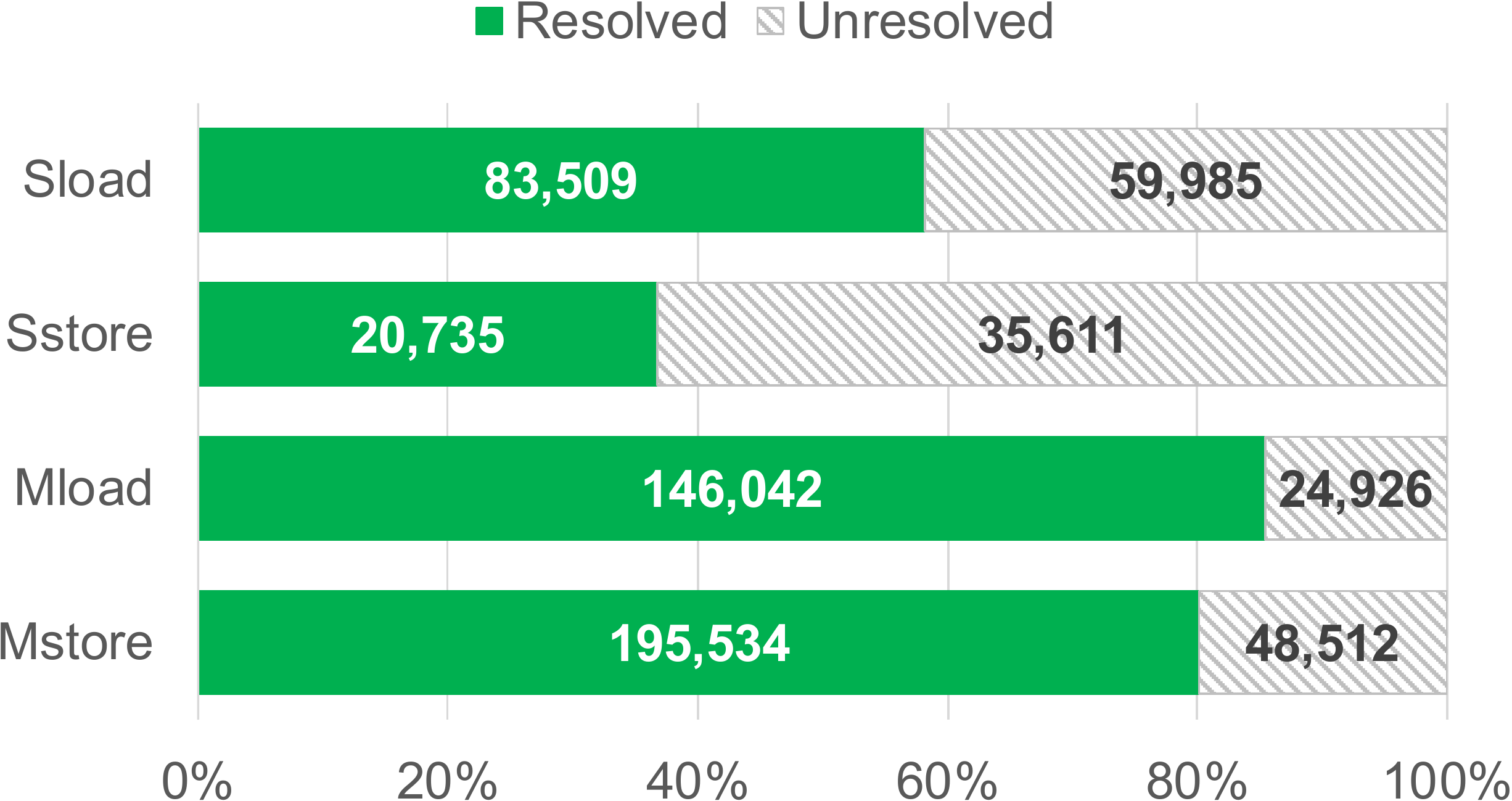}
\caption{Offsets resolved by partial evaluation.}
\label{tab:offsets}
\end{figure} 

\para{Resolving Storage/Memory Offsets}
We report on \tool's partial evaluation optimization for resolving memory and storage offsets.
Fig.~\ref{tab:offsets} shows the total number of \instruction{mload}, \instruction{mstore}, \instruction{sload} and \instruction{sstore} instructions found in our EVM dataset. The figure depicts the number of resolved offsets. On average across all four instructions, partial evaluation correctly resolves $72.6\%$ of the offsets. This indicates that \tool can often infer the precise writes to storage/memory, thereby improving the precision of the subsequent analysis. Memory offsets are more often resolved than storage offsets, as the latter often depend on user-provided inputs.


\para{Time and Memory Consumption}
\tool terminates for all contracts and takes on average $30$ seconds per contract (to check all compliance and violation patterns).
Oyente and Mythril have similar running times when used with default settings (which do not provide full coverage). To improve the coverage of these tools, users must increase the constraint solving timeouts and loop bounds, which in turn result in increased running times (especially for larger contracts). 
The memory consumption of \tool is determined by the size of the fixed point analysis.
In $95\%$ of cases, the consumption was below $10$MB, and in the rest it was below $1$GB.

\para{Summary} Overall, our results indicate that \tool's patterns are effective in finding violation and establishing correctness of contracts. Going further, we see two relevant items for future work. First, it would be interesting to integrate \tool with existing frameworks that provide formal EVM semantics, such as \cite{kevm,GrishchenkoMS18}, as a way to further validate \tool's analysis and patterns, and to formally prove the guarantees it provides. Second, we can leverage \tool to improve existing symbolic checkers, such as Oyente and Mythril. For example, \tool's compliance patterns can be used to reduce the false positive rate of these tools. 
\section{Related Work} \label{sec:related}
We discuss some of the works that are most closely related to ours.

\para{Analysis of Smart Contracts}
Smart contracts have been shown to be exposed to severe vulnerabilities~\cite{atzei2016survey,vitaly-security-blog}.
Hirai~\cite{hirai2016formal} was one of the firsts to formally verify smart contracts using the Isabelle proof assistant. In \cite{ITP-ethereum}, Hirai defines a formal model for the Ethereum Virtual Machine using the Lem language.
This model proves safety properties of smart contracts using existing interactive theorem provers.
Formal semantics of the EVM have been defined by Grishchenko~\emph{et al.}   \cite{GrishchenkoMS18} using the F* framework and by Hildenbrandt~\emph{et al.}  \cite{kevm} using the K framework~\cite{rosu-serbanuta-2010-jlap}. These semantics are executable and were validated against the official Ethereum test suite. Further, they enable the formal specification and verification of properties. 
The main benefit of these frameworks is that they provide strong formal verification guarantees and are precise (no false positives). They target arbitrary properties, but are, unfortunately, nontrivial to fully automate. In contrast, \tool targets properties that can be proved/disproved by checking simpler properties that can be verified in a fully automated way.

In the space of automated security tools for smart contracts, there are several popular systems based on symbolic execution. 
Examples include Oyente~\cite{luu2016making}, Mythril~\cite{mythril}, and Maian~\cite{maian}.
While symbolic execution is indeed a powerful generic technique for discovering bugs, it does not guarantee to explore all program paths (resulting in false negatives). 
In contrast to these tools, \tool explores all contract behaviors.
In the context of smart contracts, path constraints often involve hard-to-solve constraints, such as hash-functions, resulting in low coverage or false positives. Further, to avoid false positives, symbolic tools must precisely explore the set of feasible contract blocks.  
Towards this, Maian already uses a concrete validation step to filter false positives. An interesting application of \tool would be to filter the false positives reported by symbolic tools using its compliance patterns. 
In contrast to the approaches based on symbolic execution, 
\tool is an abstract interpreter. As such, it can provide soundness guarantees over all possible executions. This is different from 
symbolic execution which can only guarantee soundness if the number of paths can be bounded (in particular, this means that loops have to be unrolled).
 Even when the number of paths is bound, an abstract interpreter often scales better than symbolic execution since it can join paths 
  and does not have to explore different paths separately. On the other hand, symbolic execution can, in principle, handle more expressive predicates (within the logic of the underlying SMT solver), and, in theory, it has no false positives (in practice, as we show in Fig.~\ref{fig:compare}, it can have false positives).

Bhargavan \emph{et al.}~\cite{Bhargavan:2016} present preliminary work on verifying Ethereum smart contracts by translating Solidity and EVM bytecode to an existing verification system. The paper does not report how their tool performs on real-world contracts.
The work presented in~\cite{bigi2015validation} combines game theory and probabilistic model checking to validate a decentralized smart contract protocol.

The Zeus system \cite{kalra2018zeus} is a sound analyzer that translates smart contracts to the LLVM framework. Zeus uses XACML as a language to write properties. In contrast, \tool's DSL supports the checking of data- and control-flow properties. Further, Zeus does not support violation patterns as a way to reduce false positives. We could not directly compare \tool with Zeus as neither Zeus nor its benchmarks are publicly available.

Similarly to \tool, the work by Grossman~\emph{et al.}~\cite{GrossmanAGMRSZ18} also targets domains-specific properties. In more detail, they
introduce a dynamic linearizability checker to identify reentrancy issues. In contrast, \tool supports a larger class of properties for smart contracts and supports a DSL to allow security experts to extend the system with more properties.

\para{Security Factors}  
Delmolino \emph{et al.}~\cite{delmolino2016step} document the kinds of vulnerabilities students introduce while writing smart contracts and propose methods on how to avoid common pitfalls.
Chen \emph{et al.}~\cite{chen2017under} show that the current standard compiler Solidity does not properly optimize the EVM bytecode.
Seijas \emph{et al.}~\cite{seijas2016scripting} overview the capabilities of different blockchains such as Bitcoin, Nxt, and Ethereum, and survey extensions (Kosba \emph{et al.}~\cite{kosba2016hawk}).

\para{Language-Based Security}
Programming language approaches enforce security at the  program code level.
PQL~\cite{PQL05} introduces a program query language for Java that allows developers to express patterns of interest and check Java programs against them.
Both~\cite{PQL05} and our work have an underlying declarative solver for the static analysis.
Pidgin~\cite{Johnson:2015:EES:2737924.2737957} is a custom query language for program dependence graphs that can also capture security properties on Java programs.
In contrast, our work focuses on Ethereum smart contracts. \tool's analysis is tailored to the Ethereum setting, such as Ethereum-specific instructions (e.g., \instruction{balance}) and reasoning across memory and contract storage. Furthermore, \tool provides a DSL specific to security properties for smart contracts.

\para{Declarative Program Analysis}
Declarative approaches to program analysis are related to \tool's fact inference engine, as they also rely on Datalog to express analysis computations.
The Doop framework~\cite{Smaragdakis:2010:UDF:2185923.2185939,Bravenboer:2009:SDS:1640089.1640108} presents a fast and scalable declarative points-to analysis for Java programs and is one of the first works to show the promise of declarative static analysis.
Following these ideas, the authors of \cite{Zhang:2014:ARP:2666356.2594327} present a technique for automatic abstraction refinement for static analysis specified in Datalog, and in~\cite{Mangal:2015:UAP:2786805.2786851} the authors propose to involve the developer in the abstraction-refinement loop.
Researchers have developed specific extensions to Datalog, such as Flix~\cite{Madsen:2016:DFD:2908080.2908096}, to improve the efficiency and scalability of Datalog-based program analysis.
These works are orthogonal to \tool's inference engine. They develop general program analysis techniques, while \tool leverages these advances for reasoning about smart contracts.
As such, \tool can benefit from any future advances in Datalog-based program analysis. 
\section{Conclusion}\label{sec:conclusion}

We presented \tool, a new lightweight and scalable verifier for Ethereum smart contracts. \tool leverages the domain-specific insight that violations of many practical properties for smart contracts also violate simpler properties, which are significantly easier to check in a purely automated way. Based on this insight, we devised compliance and violation patterns that can effectively prove whether real-world contracts are safe/unsafe with respect to relevant properties.
Overall, \tool enjoys several important benefits:
{\em (i)}~it analyzes all contract behaviors to avoid undesirable false negatives; 
{\em (ii)}~it reduces the user effort in classifying warnings into true positives and false alarms by guaranteeing that certain behaviors are actual errors; 
{\em (iii)}~it supports a new domain-specific language that enables users to express new vulnerability patterns as they emerge; finally, 
{\em (iv)}~its analysis pipeline -- from bytecode decompilation, optimizations, to checking of patterns -- is fully automated using scalable, off-the-shelf Datalog solvers.

\section*{Acknowledgments}
The research leading to these results was partially supported by an ERC Starting Grant~$680358$. We thank Hubert Ritzdorf and the ChainSecurity team for their valuable contributions to this project.

\bibliographystyle{ACM-Reference-Format}
\bibliography{bib}

\end{document}